\documentclass[aps,prl,reprint,showpacs,superscriptaddress,longbibliography]{revtex4-1}

\usepackage{pdfpages} 
\usepackage{pgffor} 

\makeatletter
\AtBeginDocument{\let\LS@rot\@undefined}
\makeatother

\def\supplementfilename{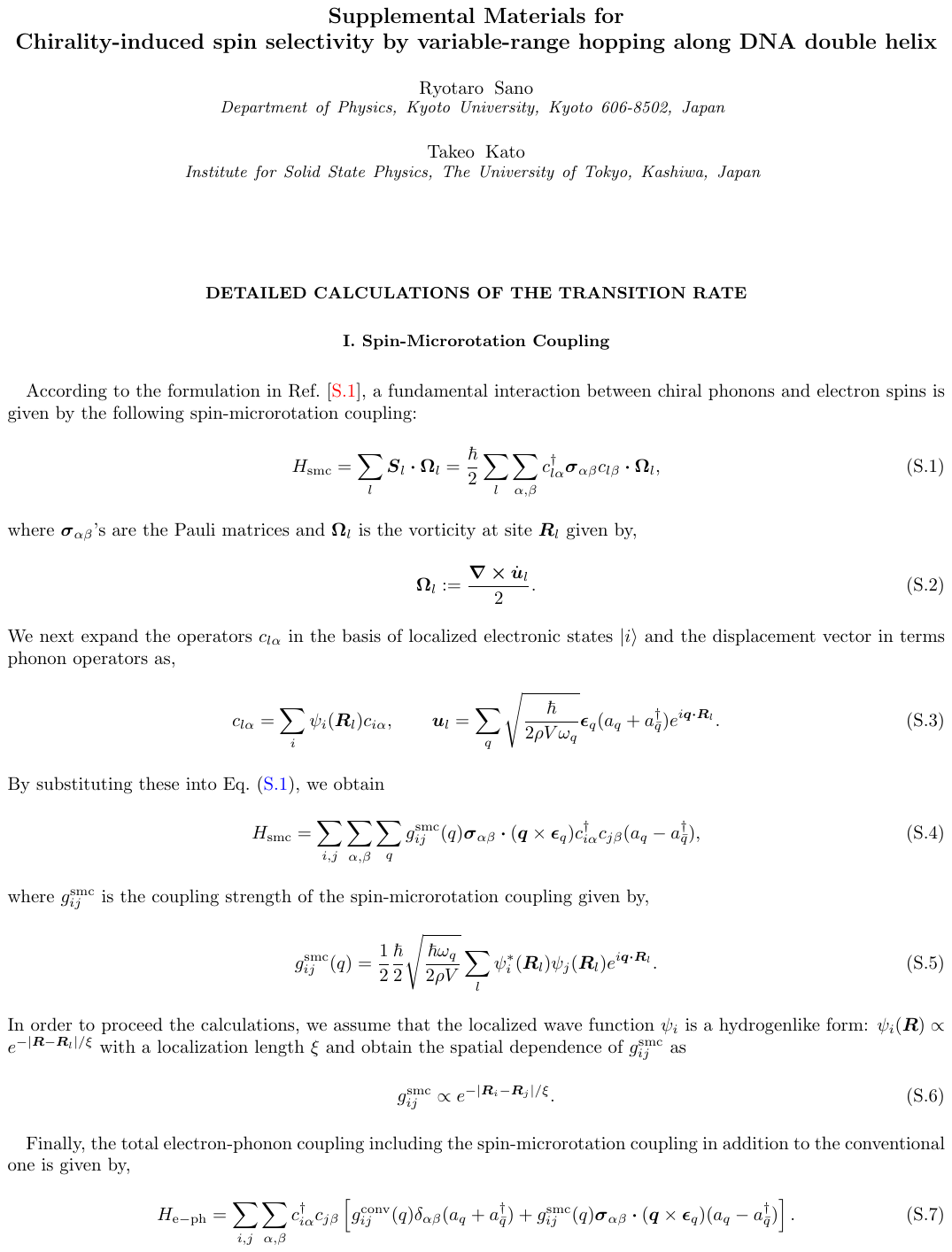}

\pdfximage{\supplementfilename}
\def\numbersupplementpages{\the\pdflastximagepages}

\newif\ifarXiv
\arXivtrue 

\usepackage{amsmath,amssymb,amsfonts}
\usepackage{comment}
\usepackage{color}
\usepackage[T1]{fontenc}
\usepackage{textcomp}
\usepackage{hyperref}
\usepackage{tikz}
\usepackage[compat=1.1.0]{tikz-feynhand}
\usepackage{multirow}
\hypersetup{
setpagesize=false,
colorlinks=true,
linkcolor=blue,
citecolor=red,
}
\usepackage{bm}
\usepackage{mathrsfs}
\usepackage{physics}
\usepackage{here}
\usepackage[normalem]{ulem}

\makeatletter
\let\MYcaption\@makecaption
\makeatother
\usepackage{subcaption}
\makeatletter
\let\@makecaption\MYcaption
\makeatother

\begin{document}
\title{Chirality-induced spin selectivity by variable-range hopping along DNA double helix}

\author{Ryotaro Sano}
\email{sano.ryotaro.52v@st.kyoto-u.ac.jp}
\affiliation{Department of Physics, Kyoto University, Kyoto 606-8502, Japan}

\author{Takeo Kato}
\affiliation{Institute for Solid State Physics, The University of Tokyo, Kashiwa, Japan}
\date{\today}

\begin{abstract}
We here present a variable-range hopping model to describe the chirality-induced spin selectivity along the DNA double helix. In this model, DNA is considered as a one-dimensional disordered system, where electrons are transported by chiral phonon-assisted hopping between localized states. Owing to the coupling between the electron spin and the vorticity of chiral phonons, electric toroidal monopole appears in the charge-to-spin conductances as a manifestation of true chirality. Our model quantitatively explains the temperature dependence of the spin polarization observed in experiments.
\end{abstract}

\maketitle

\textit{Introduction.---}
Chirality, a geometrical concept in which the structure lacks both inversion and mirror symmetries, gives a new twist to modern condensed matter physics~\cite{Barron_2004,Wagniere2007,Barron2013}. Organic molecules and structures, which are the building blocks of all organisms, commonly exhibit a well-defined chirality. The celebrated chirality-induced spin selectivity (CISS), a spin filtering effect in chiral molecules, has been extensively studied over the past few decades since its discovery in the DNA double helix~\cite{Gohler2011Science,Xie2011NanoLett}. The striking feature of CISS is an achievement of a large spin polarization even at room temperature without breaking the time-reversal symmetry~\cite{Naaman2012JPhysChemLett,Naaman2015AnnRevPhysChem,Naaman2019NatRevChem,Naaman2020JPhysChemLett,Waldeck2021APL}. Therefore, chirality is essential to emerge spin functionality in organic molecules because they consist of light elements and consequently have longer spin relaxation lengths~\cite{Rocha2005NatMater,Sanvito2011ChemSocRev,Michaeli2017JPhysCondMatt,Yang2021NatRevPhys}. The CISS effect has opened new possibilities for employing organic molecules in spintronic applications with neither the spin-orbit coupling nor magnets and for realizing enantiomer separation, which could provide a fundamental understanding on the role of electron spin in biological processes~\cite{Dor2013NatCommun,Michaeli2016ChemSocRev,BenDor2017NatCommun,Mtangi2017JAmerChemSoc,Ghosh2018Science,Tassinari2019ChemSci,Naaman2020AccChemRes,Bloom2024ChemRev}. Despite intensive efforts have been devoted to identifying the key mechanism of CISS~\cite{Yeganeh2009JChemPhys,Guo2012PRL,Kumar2013PhysChemChemPhys,Gersten2013JChemPhys,Guo2014PNAS,Kettner2015,Zwang2016JAmerChemSoc,Matityahu2016PRB,Kumar2017PNAS,Yang2019PRB,Dalum2019NanoLett,Fransson2019JPhysChemLett,Suda2019NatCommun,Mishra2020JPhysChemC,Du2020PRB,Zhang2020PRB,Fransson2020PRB,Li2020PRL,Yang2020NanoLett,Dianat2020NanoLett,Zollner2020JChemTheoCompu,Inui2020PRL,Nabei2020APL,Shiota2021PRL,Shishido2021APL,Fransson2021NanoLett,Liu2021NatMater,Wang2021NanoLett,Alwan2021JAmerChemSoc,Fransson2022JPhysChemLett,Vittmann2022JPhysChemLett,Evers2022AdvMat,Das2022JPhysChem,Clever2022IsrJChem,Qian2022Nature,Aiello2022ACSNano,Kondou2022JAmerChemSoc,Ko2022PNAS,Das2022JPhysChemC,Fransson2022IsrJChem,Kato2022PRB,Dubi2022ChemSci,Naskar2023JPhysChemLett,Fransson2023PRR,Fransson2023JChemPhys,Garcia2023JPhysChemLett,Vittmann2023JPhysChemLett,Nakajima2023Nature,Eckvahl2023Science,Kondou2023JMagMagMat,Alwan2023JChemPhys,Yang2023NatChem,Kousaka2023JJAP,Shishido2023JChemPhys,Ohe2024PRL,Sun2024NatMater,tirion2024mechanism,yan2023structural,moharana2024arXiv}, the underlying physics still remains a long-standing mystery.

The temperature dependence of the conductivity often provides rich insights into the underlying mechanisms of these transport phenomena. Experiments have demonstrated a long-range charge migration along the DNA double helix, indicating that DNA is a candidate for a one-dimensional molecular wire~\cite{Lewis1997Science,Hall1996Nature,Dandliker1997Science}. Because a DNA double helix is not a periodic system due to a random base-pair sequence, disorder effects essentially determine the electronic features of DNA. The investigation of the conductivity and its temperature dependence has revealed that electrons in DNA can be consistently described by the variable-range hopping (VRH) for the charge transport~\cite{Tran2000PRL,Yu2001PRL}. In this VRH model, the electron transport is dominated by incoherent hopping between Anderson localized states with emission or absorption of a phonon that bridges the energy difference between them, leading to the well-known Mott's law of electron transition rate~\cite{mott1997electronic,Shklovskii2013electronic,Pollak1990}.

In this Letter, we apply the Mott's VRH to the electronic spin transport along the DNA double helix. The CISS effect is commonly observed at room temperature; therefore, it is natural to expect that phonons play a crucial role. In DNA, phonons acquire chirality reflecting its chiral structure, which are the so-called chiral phonons~\cite{Zhang2015PRL,Chen2018_2DMater,Zhu2018Science,Chen2019NatPhys,Grissonnanche2020NatPhys,Chen2021NanoLett,Chen2022NanoLett,Zhang2022PRR,Yao2022PRB,Komiyama2022PRB,Skorka2023MaterTodayCommun,Tsunetsugu2023JPSJ,Kato2023JPSJ,Wang2024NanoLett}. Thus, it is essential to consider the coupling between chiral phonons and the spin degrees of freedom of electrons. This can be accomplished by the micropolar elasticity theory~\cite{Eringen_1999,eringen2001microcontinuum,Kishine2020PRL,funato2024arXiv}, which captures the rotational nature of chiral phonons~\cite{Ishito2023NatPhys,Ishito2023Chirality,Kim2023NatMater,Ueda2023Nature,Oishi2024PRB,Yao2024APL}. We then develop a framework of a random spin resistor network to describe chiral phonon-assisted hopping and numerically calculate the spin polarization based on the percolation theory~\cite{POLLAK1972JNonCrystSol,Kirkpatrick1973RevModPhys,Seager1974PRB}. The resultant temperature dependence of the spin polarization quantitatively explains observations in several experiments~\cite{Kumar2013PhysChemChemPhys,Naaman2015AnnRevPhysChem,Qian2022Nature}, indicating the relevance of both disorder effects and chiral phonons to the spin transport along the DNA double helix, which in turn will provide clues to the origin of CISS.

\textit{Formulation.---}
A DNA double helix with a random base-pair sequence can be regarded as a 1D disordered system. In this system, the disorder leads to the Anderson localization, and electron hoppings between these localized states along the chain are responsible for the conductivity. Our starting point is the following Hamiltonian:
\begin{equation}
H=H_\mathrm{e}+H_\mathrm{ph}+H_\mathrm{e-ph},
\end{equation}
where $H_\mathrm{e}=-t\sum_{l,\alpha}(c_{l\alpha}^\dagger c_{l+1\alpha}+\mathrm{h.c.})+\sum_{l,\alpha}v_lc_{l\alpha}^\dagger c_{l\alpha}$ describes electrons hopping with amplitude $t$ on a 1D lattice with on-site random potential $v_l$ at site $\vb*{R}_l$. $v_l$ is uniformly distributed in the interval $[-W,W]$, where $W$ is the strength of the disorder. We can rewrite $H_\mathrm{e}=\sum_{i,\alpha}\varepsilon_ic_{i\alpha}^\dagger c_{i\alpha}$ in the basis of localized electronic states $\ket{i}$, i.e., $\psi_i(x)\sim e^{-\abs{x-x_i}/\xi}/\sqrt{\xi}$ with energy $\varepsilon_i$ and a localization length $\xi$.

The part $H_\mathrm{ph}$ describes the chiral phonons reside at the molecular wire:
\begin{equation}
    H_\mathrm{ph}=\sum_l\left[\frac{\vb*{p}_l^2}{2M}+\frac{K}{2}(\vb*{u}_l-\vb*{u}_{l+1})^2\right]=\sum_q\hbar\omega_q a_q^\dagger a_q,
\end{equation}
where $a_{\bar{q}}^\dagger$ and $a_q$ are the phonon creation and annihilation operators with $q=(\vb*{q},\lambda)$ [$\bar{q}=(-\vb*{q},\bar{\lambda})$], which are related to the displacement vector $\vb*{u}(\vb*{r})=\sum_q\sqrt{\frac{\hbar}{2\rho V\omega_q}}\vb*{\epsilon}_q (a_q+a^\dagger_{\bar{q}})e^{\mathrm{i}\vb*{q}\vdot\vb*{r}}$. Here, $\rho$ is the mass density, $V$ is the volume of the system, $\omega_q$ is the phonon dispersion, and $\vb*{\epsilon}_q$ is the displacement polarization vector, which satisfies the orthonormal condition: $\vb*{\epsilon}_{\vb*{q}\lambda}^\ast\vdot\vb*{\epsilon}_{\vb*{q}\lambda'}=\delta_{\lambda\lambda'}$. 
Due to the broken inversion and mirror symmetries with preserving the time-reversal symmetry, which is the modern definition of true chirality~\cite{Barron_2004,Barron2013}, $\omega_{{q}}=\omega_{\bar{q}}\neq\omega_{-\vb*{q}\lambda}$ and $\vb*{\epsilon}^\ast_{q}=\vb*{\epsilon}_{\bar{q}}\neq\vb*{\epsilon}_{-\vb*{q}\lambda}$ hold for chiral phonons~\cite{Kishine2020PRL}.
Note that the structural and the dynamical chiralities are encoded in $\omega_q$ and $\vb*{\epsilon}_q$.

\begin{figure}[t]
\centering
  \begin{tikzpicture}
      \begin{feynhand}
      \vertex[dot](a_1)at(-3.5,0);
      \vertex[dot](a_2)at(-1.5,0){};
      \vertex[dot](a_3)at(1.5,0){};
      \vertex[dot](a_4)at(3.5,0);
      \propag[chasca](a_1)to[edge label'={$i,\alpha,\mathrm{i}\omega_n$}](a_2);
      \propag[fer](a_2)to[edge label'={$j,\beta,\mathrm{i}\omega_n-\mathrm{i}\nu_\ell$}](a_3);
      \propag[chasca](a_3)to[edge label'={$i,\alpha,\mathrm{i}\omega_n$}](a_4);
      \propag[photon](a_2)to[in = 90,out = 90, edge label = {$q,\mathrm{i}\nu_\ell$}](a_3);
    \end{feynhand}
    \end{tikzpicture}\\
  \caption{Second-order self-energy diagram considered for estimating the hopping rate from a localized state $(i,\alpha)$ to another localized state $(j,\beta)$.}
  \label{fig:self-energy}
\end{figure}
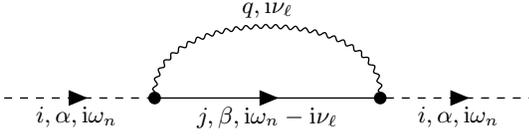

For organic molecules which consist of light elements with negligible spin-orbit interactions, the conventional electron-phonon coupling does not affect the spin degrees of freedom of electrons and thus cannot trigger the CISS effect. To overcome this difficulty, we go beyond the conventional elasticity framework and employ the micropolar elasticity theory~\cite{Eringen_1999,eringen2001microcontinuum}, which captures the rotational nature of chiral phonons~\cite{Kishine2020PRL}. Then, the total electron-phonon coupling $H_\mathrm{e-ph}$ includes both the conventional type of lattice deformation originating from a lattice displacement~\cite{mahan} and the novel type originating from a vorticity, which is the adiabatic limit of a microrotation~\cite{Kishine2020PRL}. We here focus on the latter one given by,
\begin{align}
H_\mathrm{smc}&=\sum_{i,j}\sum_{\alpha,\beta}\sum_{q}g_{ij}^\mathrm{smc}(q)\vb*{\sigma}_{\alpha\beta}\vdot(\vb*{q}\times\vb*{\epsilon}_q)c_{i\alpha}^\dagger c_{j\beta}(a_q-a_{\bar{q}}^\dagger).
\end{align}
The polarization vector $\vb*{\epsilon}_q$ carries the dynamical chirality of chiral phonons; therefore, enables a coupling between the vorticity of chiral phonons and the electron spin dubbed the spin-microrotation coupling (SMC)~\cite{funato2024arXiv}. $\vb*{\sigma}_{\alpha\beta}$'s are the Pauli matrices for the electron spin and $g_{ij}^\mathrm{smc}(q)=[g_{ji}^{\mathrm{smc}}(\bar{q})]^\ast$ is the coupling strength of SMC. Note that only transverse components of chiral phonons contribute to SMC. Therefore, this SMC can be regarded as a fundamental interaction between chiral phonons and electron spins. Here and hereafter, we assume that the global spin quantization axis is chosen along the chiral axis of the DNA double helix.

In the weak coupling approach, we evaluate the hopping rate assisted by single phonon processes using perturbation theory in $H_\mathrm{smc}$. By considering the self-energy diagram of Fig.~\ref{fig:self-energy} that describes processes through which a localized state $(i,\alpha)$ decay to other localized states $(j,\beta)$ by emitting or absorbing a chiral phonon, the second-order perturbation for SMC leads to Fermi's golden rule,
\begin{widetext}
\begin{align}
    \frac{1}{\tau_{i\alpha}^{\mathrm{smc}}}&=\frac{2\pi}{\hbar}\sum_{j,\beta}\sum_q\abs{g_{ij}^\mathrm{smc}(q)[\vb*{\sigma}_{\alpha\beta}\vdot(\vb*{q}\times\vb*{\epsilon}_q)]}^2[\{1-f(\varepsilon_{j})+n(\hbar\omega_q)\}\delta(\varepsilon_i-\varepsilon_{j}-\hbar\omega_q)+\{f(\varepsilon_{j})+n(\hbar\omega_q)\}\delta(\varepsilon_i-\varepsilon_{j}+\hbar\omega_q)]\nonumber\\
    &=\sum_{j,\beta}\Gamma^0_{(i\alpha)\to(j\beta)},\label{smc:transition-rate}
\end{align}
\end{widetext}
where $f(\varepsilon)$ and $n(\hbar\omega)$ are the Fermi and the Bose distribution functions, respectively. $\delta(\varepsilon)$ is a delta function in energy. Here, $\Gamma^0_{(i\alpha)\to(j\beta)}$ describes the transition rate from a state $(i,\alpha)$ to another state $(j,\beta)$. It is worth noting that SMC gives rise to emergent terms in $\tau_{i\alpha}^{-1}$:
\begin{align}
&\sum_{\beta}\abs{\vb*{\sigma}_{\alpha\beta}\vdot(\vb*{q}\times\vb*{\epsilon}_q)}^2\nonumber\\
&=\delta_{\alpha\alpha}[\vb*{q}^2-(\vb*{q}\vdot\vb*{\epsilon}_q)(\vb*{q}\vdot\vb*{\epsilon}_q^\ast)]+(\vb*{q}\vdot\vb*{\sigma}_{\alpha\alpha})[\vb*{q}\vdot\Im(\vb*{\epsilon}_q^\ast\times\vb*{\epsilon}_q)].\label{G0}
\end{align} 
The two factors, $(\vb*{q}\vdot\vb*{\sigma}_{\alpha\alpha})$ and $[\vb*{q}\vdot\mathrm{Im}(\vb*{\epsilon}_q^\ast\times\vb*{\epsilon}_q)]$, are the inner products of a time-reversal ($\mathcal{T}$)-odd polar and a $\mathcal{T}$-odd axial vector (refer to Table~\ref{table:symmetry}).
Therefore, these factors are both $\mathcal{T}$-even pseudoscalars and belong to the electric toroidal monopole $G_0$~\cite{Hayami2018PRB}, which manifests true chirality of materials~\cite{Kishine2022Israel}.
Remarkably, the product of these two factors is a unique combination to construct a scalar quantity $\tau_{i\alpha}^{-1}$ by pseudoscalars associated with both electron spins and chiral phonons.
Since $[\vb*{q}\vdot\mathrm{Im}(\vb*{\epsilon}_q^\ast\times\vb*{\epsilon}_q)]$ becomes nonzero when phonons exhibit chirality, the transition rate can acquire a spin dependence through the factor $(\vb*{q}\vdot\vb*{\sigma}_{\alpha\alpha})$ by this type of electron-phonon coupling.
The diagonal elements of the Pauli matrices survive only for the component in the spin quantization direction, which we take along the chiral axis of the system. Therefore, the preferred spin orientation is parallel or antiparallel to the axis, which direction depends on true chirality of the material.

 \begin{table}[t]
     \centering
     \caption{Symmetry classification and comparison of the physical quantities appears in the transition rate under the inversion $\mathcal{P}$ and the time-reversal $\mathcal{T}$ operations. The symbols of multipoles follow the literature~\cite{Hayami2018PRB}.}
     \begin{tabular}{c|ccc|c|c}
     \hline
      & $\mathcal{P}$ & & $\mathcal{T}$ & type & Multipole \\
     \hline \hline
     $\vb*{q}$ & $-$ & & $-$ & \multirow{2}{*}{polar} & $T_{1m}$\\
$\vb*{\epsilon}_q$ & $-$ & & $+$ &  & $Q_{1m}$\\
\hline
     $\vb*{\sigma}_{\alpha\beta}$ & $+$ & & $-$ & \multirow{3}{*}{axial} & \multirow{3}{*}{$M_{1m}$}\\
     $\vb*{q}\times\vb*{\epsilon}_q$ & $+$ & & $-$ &  & \\
$\mathrm{Im}(\vb*{\epsilon}_q^\ast\times\vb*{\epsilon}_q)$ & $+$ & & $-$ &  & \\
\hline
$\mathrm{i}\vb*{q}\vdot\vb*{\epsilon}_q$ & $+$ & & $+$ &\multirow{2}{*}{scalar} & \multirow{2}{*}{$Q_0$}\\
$\vb*{\sigma}_{\alpha\beta}\vdot(\vb*{q}\times\vb*{\epsilon}_q)$ & $+$ & & $+$ &  & \\
\hline
     $\vb*{\sigma}_{\alpha\beta}\vdot\vb*{q}$ & $-$ & & $+$ & \multirow{2}{*}{pseudo scalar} & \multirow{2}{*}{$G_0$}\\
     $\vb*{q}\vdot\mathrm{Im}(\vb*{\epsilon}_q^\ast\times\vb*{\epsilon}_q)$ & $-$ & & $+$ &  & \\
     \hline 
     \end{tabular}
     \label{table:symmetry}
 \end{table}

\textit{Variable-range hopping.---}
Before applying the above microscopic calculations to CISS in the DNA double helix, we introduce the conventional variable-range hopping (VRH) scheme~\cite{mott1997electronic,Shklovskii2013electronic,Pollak1990}. VRH is a model describing low-temperature conduction in strongly disordered systems with localized states. To understand the VRH transport, the first ingredient is the transition rates between localized states due to the electron-phonon coupling. The incoherent hopping conduction is a result of many series of such transitions. In a phonon-assisted hopping process, an electron is transferred from a single-particle localized state $\ket{i}$ centered at site $x_i$ to the localized state $\ket{j}$ together with emission or absorption of a phonon that bridges the energy difference $\abs{\varepsilon_i-\varepsilon_j}$ between the two states. The transition matrix element is proportional to the spatial overlap of the electronic states and hence decay exponentially with the range of the hop as $e^{-2\abs{x_i-x_j}/\xi}$.

The hopping rate out of a localized electronic state can be obtained through Fermi's golden rule as~\cite{Yu2001PRL,Sumilan2016PRL}
\begin{equation}
    \frac{1}{\tau_\mathrm{VRH}}\simeq g^2\sum_Re^{-2R/\xi}e^{-\Delta_R/2k_\mathrm{B}T}\nu(\Delta_R),\label{VRH:phenomenon}
\end{equation}
where $\Delta_R\simeq\Delta_\xi(\xi/R)^d$ is the typical energy offset to the nearest state localized within a range $R$ of the initial state, which is supplied by a thermal phonon, and $\Delta_\xi=2W(a/\xi)^d$ is the average energy level spacing within a localization volume $\xi^d$. $\nu(\varepsilon)$ is the density of states for phonons and $g$ is the strength of the electron-phonon coupling. There is a competition between the terms in the exponential and an electron may optimize its hopping distance to achieve the largest hopping rate. Then, the hopping range at a given temperature $R_\mathrm{opt}=\xi(\Delta_\xi/4k_\mathrm{B}T)^{1/(d+1)}$ can be obtained from the saddle point of the sum in Eq.~\eqref{VRH:phenomenon}. Finally, replacing the sum with the saddle point values gives the well-known Mott's law for the electron hopping rate: $\tau_\mathrm{opt}^{-1}\sim\exp[-(T_0/T)^{1/(d+1)}]$, where $k_\mathrm{B}T_0=2W(4a/\xi)^{d}$.

\begin{figure}[t]
\includegraphics[keepaspectratio,scale=0.34]{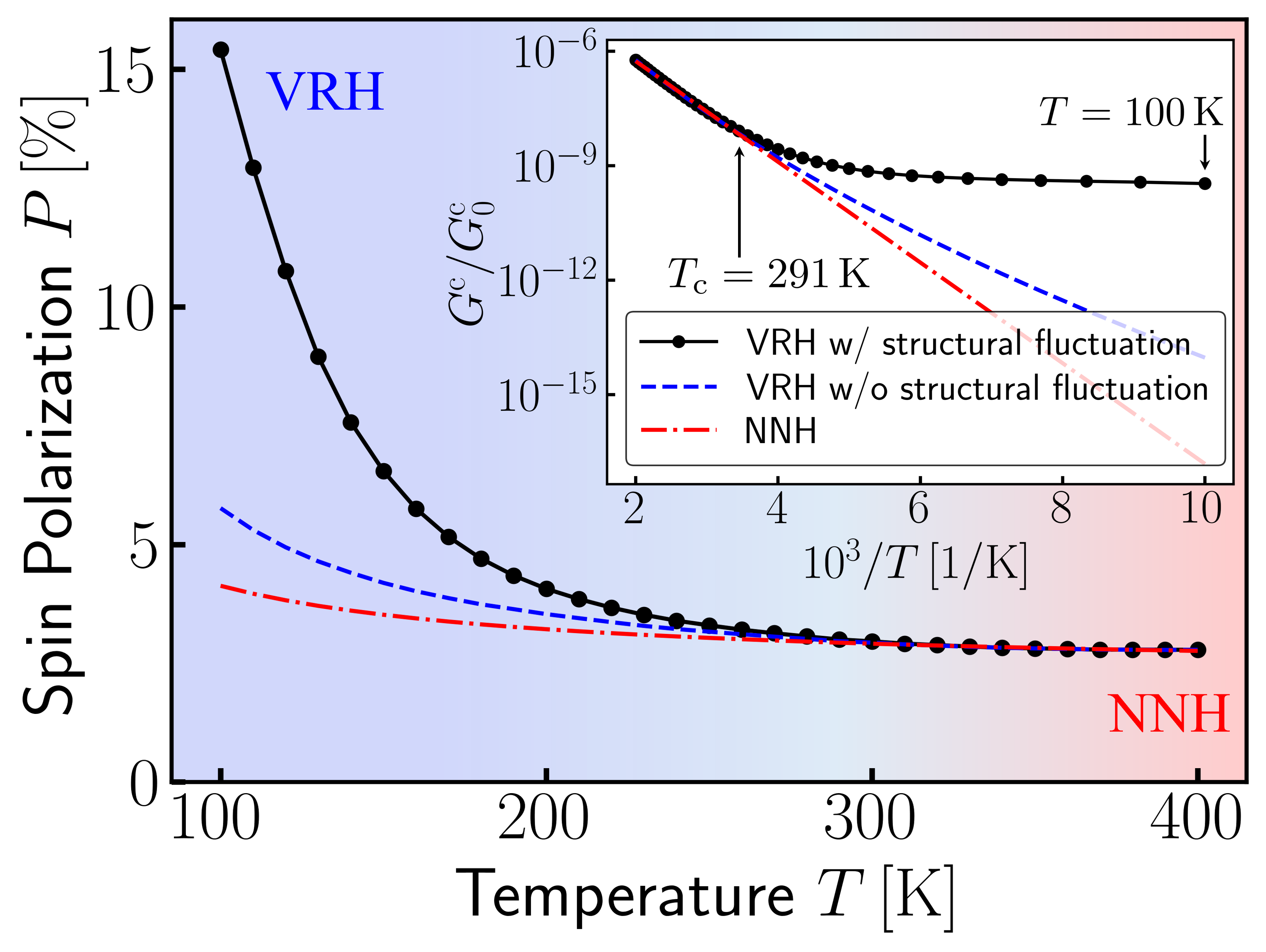}
  \caption{Temperature dependence of the spin polarization $P$ calculated by a random spin resistor network ranging from $100$ to $400$\,K with a molecular length of 40 base-pairs. The inset depicts the charge conductance $G^\mathrm{c}$ as a function of inverse temperature $1/T$ and shows that $T_\mathrm{c}=291\,\mathrm{K}$ is the crossover temperature between the nearest-neighbor hopping (NNH) and the variable-range hopping (VRH). Here, we have used the parameters: $t = 0.065$\,eV, $W = 0.15$\,eV, and $a = 3.4$\,\AA. We have also included the temperature dependence of the localization length: $\xi^{-1}(T) = 0.18+0.70\tanh(T/192)^2$\,\AA${}^{-1}$, which originates from thermal structural fluctuations~\cite{Yu2001PRL}.}
  \label{fig:spin-RRN}
\end{figure}

The criterion, $R_\mathrm{opt}=a$, naturally gives the crossover temperature: $k_\mathrm{B}T_\mathrm{c}=W\xi/2a$. At high temperatures $T>T_\mathrm{c}$, electron transport is via the nearest-neighbor hopping (NNH) and is a simple thermal activation process, whereas at low temperatures $T<T_\mathrm{c}$, electrons optimize their paths via the VRH mechanism and are more likely to jump to a remote site [see the inset of Fig.~\ref{fig:spin-RRN}].

\textit{Random spin resistor network.---}
We are now ready to discuss the electronic spin transport by the VRH mechanism.
Corresponding to many transport experiments of CISS measurements~\cite{Xie2011NanoLett}, we here consider the situation where an electric field is applied along the chiral axis of the system, driving it into nonequilibrium.
In the linear response regime to the electric field, the steady-state conductance can be calculated from the percolation theory~\cite{POLLAK1972JNonCrystSol,Kirkpatrick1973RevModPhys,Seager1974PRB}. Then, the problem can be mapped into an equivalent random resistor network~\cite{Miller1960PR,Ambegaokar1971PRB,Lee1984PRL}, where each pair of sites is connected by a resistance related to the corresponding transition rate. We here extend this method to incorporate the spin transport while satisfying the charge conservation, namely, the Kirchhoff's law. 

The net charge current from a site $i$ with spin $\alpha$ to a site $j$ with spin $\beta$ is given by,
\begin{equation}
I_{ij}^{\alpha\beta}=G_{ij}^{\alpha\beta}(V_i^\alpha-V_j^\beta),\label{spin-dependent-conductance}
\end{equation}
where $G_{ij}^{\alpha\beta}=\frac{e^2}{k_\mathrm{B}T}\Gamma^0_{(i\alpha)\to(j\beta)}f(\varepsilon_i)$ and $-eV_i^\alpha=e\vb*{E}\vdot\vb*{R}_i+\delta\mu_i^\alpha$ are the conductance and the spin-dependent electrochemical potential, respectively. Here, $\delta\mu_i^\alpha$ is the nonequilibrium chemical potential at site $i$ with spin $\alpha$. In contrast to the previous studies on a random resistor network, we should reside two spin states at each site.

We next introduce the generalized Kirchhoff's law~\cite{Hamrle2005JApplPhys,Camsari2015SciRep,Camsari2019IEEE,Baez2020PRB} and transform it into the charge-spin basis to correspond with experimental conditions:
\begin{subequations}
\begin{equation}
    \sum_{j\in Z(i)}{\vb{G}}_{ij}{\vb{V}}_j+{\vb{G}}_{ii}{\vb{V}}_i+({\vb{I}}_i)^\mathrm{source}=0,
\end{equation}
where the sum runs over the set of all sites $Z(i)$ connected to the site $i$. Here, we have defined the conductance matrices in the charge-spin basis as
    \begin{equation}
        {\vb{G}}_{ij}=\left[
  \begin{array}{cc}
    {G}_{ij}^\mathrm{cc} & {G}_{ij}^\mathrm{cs}\\
    {G}_{ij}^\mathrm{sc} & {G}_{ij}^\mathrm{ss}
  \end{array}
  \right],\ \ 
{\vb{G}}_{ii}=-\sum_{j\in Z(i)}\left[
  \begin{array}{cc}
      {G}_{ij}^{\mathrm{cc}} & {G}_{ij}^{\mathrm{sc}}\\
 {G}_{ij}^{\mathrm{sc}} & {G}_{ij}^{\mathrm{cc}}
  \end{array}
  \right],
    \end{equation}
and the charge/spin voltages and currents as
    \begin{equation}
        {\vb{V}}_i=\left[
  \begin{array}{c}
    V_i^\mathrm{c}\\
    V_i^\mathrm{s}
  \end{array}
  \right]=\frac{1}{2}\left[
  \begin{array}{c}
    V_i^\uparrow+V_i^\downarrow\\
    V_i^\uparrow-V_i^\downarrow
  \end{array}
  \right],\ \
  {\vb{I}}_i=\left[
\begin{array}{c}
  I_i^\mathrm{c}\\
  I_i^\mathrm{s}
\end{array}
\right]=\left[
\begin{array}{c}
  I_i^\uparrow+I_i^\downarrow\\
  I_i^\uparrow-I_i^\downarrow
\end{array}
\right],
    \end{equation}
\end{subequations}
respectively. Note that the symmetry $G_{ij}^{\alpha\beta}=G_{ji}^{\beta\alpha}$ in Eq.~\eqref{spin-dependent-conductance} guarantees the absence of both charge and spin currents in equilibrium (with no bias voltage)~\footnote{See the Supplemental Materials for detailed derivations of the conductance matrices and the sum rules.}. Each component of $\vb{G}_{ij}$ is given by 
\begin{subequations}\label{conductance:charge/spin} 
\begin{align}
{G}_{ij}^\mathrm{cc}&=G_{ij}^{\uparrow\uparrow}+G_{ij}^{\uparrow\downarrow}+G_{ij}^{\downarrow\uparrow}+ G_{ij}^{\downarrow\downarrow},\\
G_{ij}^\mathrm{cs}&=G_{ij}^{\uparrow\uparrow}+G_{ij}^{\downarrow\uparrow}-G_{ij}^{\uparrow\downarrow}-G_{ij}^{\downarrow\downarrow},\\
G_{ij}^\mathrm{sc}&=G_{ij}^{\uparrow\uparrow}+G_{ij}^{\uparrow\downarrow}-G_{ij}^{\downarrow\uparrow}-G_{ij}^{\downarrow\downarrow},\\
G_{ij}^\mathrm{ss}&=G_{ij}^{\uparrow\uparrow}+G_{ij}^{\downarrow\downarrow}-G_{ij}^{\uparrow\downarrow}-G_{ij}^{\downarrow\uparrow},
\end{align}
\end{subequations}
and represents how much a charge or a spin current is generated by a charge or a spin voltage. Our theory spontaneously includes driving forces stemming from the spin accumulation $\mu^\mathrm{s}_i=-eV_i^\mathrm{s}$ in addition to the conventional bias voltage $V_i^\mathrm{c}$. As can be seen from the above definitions in Eq.~\eqref{conductance:charge/spin}, $G_{ij}^\mathrm{cs}$ and $G_{ij}^\mathrm{sc}$, which activate mutual conversion between charge and spin, have the form of $\sum_{\beta}G_{ij}^{\alpha\beta}$ and hence exhibit true chirality [see Eq.~\eqref{G0}]. Therefore, the interplay between the structural and the dynamical chiralities serves as the charge-spin converter in the conductance expression. Note that the imbalance between $G_{ij}^\mathrm{\uparrow\uparrow}$ and $G_{ij}^{\downarrow\downarrow}$ is essential to realize a finite spin polarization, which needs not only the spin-microrotation coupling but also the conventional one~\footnote{See the Supplemental Materials for the detailed calculations of electron transition rates including the conventional electron-phonon coupling.}.

\textit{Application to the DNA double helix.---}
To quantitatively study the temperature dependence of the charge conductance and the spin polarization along the DNA double helix, we numerically calculate them by applying a random spin resistor network with the simplified form of conductances:
\begin{equation}
    G_{ij}^{\alpha\beta}=G_0^{\alpha\beta}\exp[-\frac{2\abs{x_i-x_j}}{\xi}-\frac{\abs{\varepsilon_i}+\abs{\varepsilon_j}+\abs{\varepsilon_i-\varepsilon_j}}{2k_\mathrm{B}T}],\label{G:percolation}
\end{equation}
where $G^{\alpha\beta}_0$ is the spin-dependent prefactor. The most important parameters in Eq.~\eqref{G:percolation} are the random variables $\varepsilon_i$, $\varepsilon_j$, and $\abs{x_i-x_j}$.
Due to the exponential spread in the values of the resistors, the conductance of the entire network will be determined by the largest conductance such that the network, which is composed of all resistors with conductances larger than a critical value, percolates.

Each spin component accumulates in the same amount at the opposite ends of the system~\footnote{See the Supplemental Materials for spatial profiles of the spin accumulation\label{supple:sinh}}. Then, we here define the spin polarization $P$ as the difference of the spin accumulation at the two ends,
\begin{equation}
    P=\frac{\mu_1^\mathrm{s}-\mu_\mathrm{N}^\mathrm{s}}{eV}=-\frac{V_1^\mathrm{s}-V_\mathrm{N}^\mathrm{s}}{V}.
\end{equation}
The temperature dependence of $P$ is shown in Fig.~\ref{fig:spin-RRN}. As decreasing temperature, the underlying mechanism of electron transport changes from NNH to VRH, which gives rise to an enhancement of the spin polarization governed by a universal power law: $P\propto 1/T^{3/2}$~\footnote{See the Supplemental Materials for fitting results for various parameters.}. Fig.~\ref{fig:spin-RRN} is the main result of this Letter and is in good agreement with the experimental observations~\cite{Kumar2013PhysChemChemPhys,Naaman2015AnnRevPhysChem,Qian2022Nature}. 

\textit{Universal $1/T^{3/2}$-law.---}
We have clarified that the spatial profile of the spin accumulation is well described by a steady-state diffusion equation $l_\mathrm{sd}^2\partial_x^2\mu^\mathrm{s}(x)=\mu^\mathrm{s}(x)$ with $\sinh[(x-L/2)/l_\mathrm{sd}]$ as a solution, where $L=\mathrm{N}a$ and $l_\mathrm{sd}\propto T(a/R_\mathrm{opt})$ are the system and the spin diffusion lengths~\cite{Note3}. Thus, we have identified the origin of the temperature dependence of $P$, which is proportional to $2\sinh[L/2l_\mathrm{sd}]\sim l_\mathrm{sd}^{-1}$, as that of the optimized hopping range $R_\mathrm{opt}$ associated with the crossover from NNH to VRH. This gives rise to a universal $1/T^{3/2}$-law for $d=1$ in the VRH regime.

Finally, we should note that the amplitude of the spin polarization and its sign depend on the structural and the dynamical chiralities, the electron-phonon coupling strengths, and other material parameters, whereas its temperature dependence relies only on the underlying mechanism of electron transport. Therefore, we conclude that chiral phonon-assisted hopping between localized states is a key physics of the CISS effect along the DNA double helix.

\textit{Conclusion.---}
In summary, we have proposed a model to describe the electron charge and spin transport along the DNA double helix, where DNA is regarded as a 1D disordered system and chiral phonon-assisted hopping between localized states is the main mechanism. By employing the second-order perturbation for the spin-microrotation coupling, we have elucidated that the charge-to-spin conductances are described by the electric toroidal monopole which is a manifestation of true chirality.
In order to conduct numerical calculations, we have developed a framework of a random spin resister network, which enables the observation of the crossover of the underlying physics from the nearest-neighbor hopping to the variable-range hopping as decreasing temperature. Our results quantitatively agree with spin polarization measurements in DNA and indicates that the variable-range hoping may be crucial to the understanding of the CISS effect along the DNA double helix.
Therefore, our results give an insight on the relevance of both disorder effects and chiral phonons to CISS and will motivate further research on the temperature dependence of the spin polarization, which in turn lead to the solution of the long-standing mystery of its origin.

\begin{acknowledgements}
The authors are grateful to S. Kaneshiro, A. Kofuji, S. Asano, K. Nogaki, J. Tei, Y. Suzuki, T. Sato, T. Funato, M. Matsuo, and J. Kishine for valuable discussions. R.S. thanks M. Komoike for providing helpful comments from an experimental point of view. R.S. is supported by JSPS KAKENHI (Grants JP 22KJ1937). 
T.K. is supported by JSPS KAKENHI (Grants JP 24K06951).
\end{acknowledgements}
\bibliography{CISS.bib}

\ifarXiv
    

\section*{Supplementary material}

# Supplemental Materials for
# Chirality-induced spin selectivity by variable-range hopping along DNA double helix

Ryotaro Sano
*Department of Physics, Kyoto University, Kyoto 606-8502, Japan*

Takeo Kato
*Institute for Solid State Physics, The University of Tokyo, Kashiwa, Japan*

## DETAILED CALCULATIONS OF THE TRANSITION RATE

### I. Spin-Microrotation Coupling

According to the formulation in Ref. [S.1], a fundamental interaction between chiral phonons and electron spins is given by the following spin-microrotation coupling:

$$H_{\rm smc} = \sum_l \bm{S}_l \cdot \bm{\Omega}_l = \frac{\hbar}{2} \sum_l \sum_{\alpha,\beta} c^\dagger_{l\alpha} \bm{\sigma}_{\alpha\beta} c_{l\beta} \cdot \bm{\Omega}_l, \tag{S.1}$$

where $\bm{\sigma}_{\alpha\beta}$'s are the Pauli matrices and $\bm{\Omega}_l$ is the vorticity at site $\bm{R}_l$ given by,

$$\bm{\Omega}_l := \frac{\nabla \times \dot{\bm{u}}_l}{2}. \tag{S.2}$$

We next expand the operators $c_{l\alpha}$ in the basis of localized electronic states $|i\rangle$ and the displacement vector in terms phonon operators as,

$$c_{l\alpha} = \sum_i \psi_i(\bm{R}_l) c_{i\alpha}, \qquad \bm{u}_l = \sum_q \sqrt{\frac{\hbar}{2\rho V \omega_q}} \bm{\epsilon}_q (a_q + a^\dagger_{\bar{q}}) e^{i\bm{q}\cdot\bm{R}_l}. \tag{S.3}$$

By substituting these into Eq. (S.1), we obtain

$$H_{\rm smc} = \sum_{i,j} \sum_{\alpha,\beta} \sum_q g^{\rm smc}_{ij}(q) \bm{\sigma}_{\alpha\beta} \cdot (\bm{q} \times \bm{\epsilon}_q) c^\dagger_{i\alpha} c_{j\beta} (a_q - a^\dagger_{\bar{q}}), \tag{S.4}$$

where $g^{\rm smc}_{ij}$ is the coupling strength of the spin-microrotation coupling given by,

$$g^{\rm smc}_{ij}(q) = \frac{1}{2} \frac{\hbar}{2} \sqrt{\frac{\hbar \omega_q}{2\rho V}} \sum_l \psi^*_i(\bm{R}_l) \psi_j(\bm{R}_l) e^{i\bm{q}\cdot\bm{R}_l}. \tag{S.5}$$

In order to proceed the calculations, we assume that the localized wave function $\psi_i$ is a hydrogenlike form: $\psi_i(\bm{R}) \propto e^{-|\bm{R}-\bm{R}_l|/\xi}$ with a localization length $\xi$ and obtain the spatial dependence of $g^{\rm smc}_{ij}$ as

$$g^{\rm smc}_{ij} \propto e^{-|\bm{R}_i - \bm{R}_j|/\xi}. \tag{S.6}$$

Finally, the total electron-phonon coupling including the spin-microrotation coupling in addition to the conventional one is given by,

$$H_{\rm e-ph} = \sum_{i,j} \sum_{\alpha,\beta} c^\dagger_{i\alpha} c_{j\beta} \left[ g^{\rm conv}_{ij}(q) \delta_{\alpha\beta} (a_q + a^\dagger_{\bar{q}}) + g^{\rm smc}_{ij}(q) \bm{\sigma}_{\alpha\beta} \cdot (\bm{q} \times \bm{\epsilon}_q)(a_q - a^\dagger_{\bar{q}}) \right]. \tag{S.7}$$

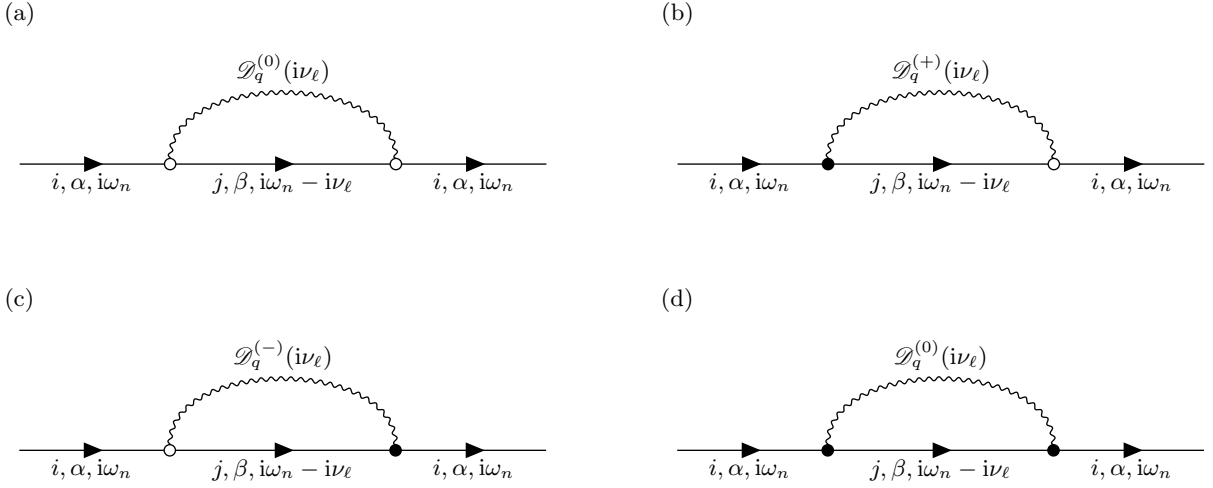

Supplementary Figure S1. Second-order self-energy diagrams. The solid line with arrow corresponds to electron propagator and the curly line to the phonon propagator. At the vertices, the coupling strength of the electron-phonon interaction should be associated. Here, the white circle ($\circ$) corresponds to $g_{ij}^{\text{conv}}(q)\delta_{\alpha\beta}(i\boldsymbol{q}\cdot\boldsymbol{\epsilon}_q)$ and the black circle ($\bullet$) to $g_{ij}^{\text{smc}}(q)\boldsymbol{\sigma}_{\alpha\beta}\cdot(\boldsymbol{q}\times\boldsymbol{\epsilon}_q)$.

## II. Perturbative Hopping Rate

The single-phonon process in Fig. S1 gives rise to the second-order self-energy,

$$\Sigma_{\alpha\alpha}^{(a)}(i,i;i\omega_n) = -\frac{1}{\beta}\sum_{j,\beta}\sum_q g_{ij}^{\text{conv}}(q)g_{ji}^{\text{conv}}(\bar{q})\delta_{\alpha\beta}\sum_{i\nu_\ell}\mathscr{G}_j^{(0)}(i\omega_n - i\nu_\ell)\mathscr{D}_q^{(0)}(i\nu_\ell), \tag{S.8a}$$

$$\Sigma_{\alpha\alpha}^{(b)}(i,i;i\omega_n) = -\frac{1}{\beta}\sum_{j,\beta}\sum_q g_{ij}^{\text{smc}}(q)g_{ji}^{\text{conv}}(\bar{q})\boldsymbol{\sigma}_{\alpha\beta}\cdot(\boldsymbol{q}\times\boldsymbol{\epsilon}_q)\delta_{\alpha\beta}\sum_{i\nu_\ell}\mathscr{G}_j^{(0)}(i\omega_n - i\nu_\ell)\mathscr{D}_q^{(+)}(i\nu_\ell), \tag{S.8b}$$

$$\Sigma_{\alpha\alpha}^{(c)}(i,i;i\omega_n) = -\frac{1}{\beta}\sum_{j,\beta}\sum_q g_{ij}^{\text{conv}}(q)g_{ji}^{\text{smc}}(\bar{q})\boldsymbol{\sigma}_{\alpha\beta}\cdot(-\boldsymbol{q}\times\boldsymbol{\epsilon}_q)\delta_{\alpha\beta}\sum_{i\nu_\ell}\mathscr{G}_j^{(0)}(i\omega_n - i\nu_\ell)\mathscr{D}_q^{(-)}(i\nu_\ell), \tag{S.8c}$$

$$\Sigma_{\alpha\alpha}^{(d)}(i,i;i\omega_n) = -\frac{1}{\beta}\sum_{j,\beta}\sum_q g_{ij}^{\text{smc}}(q)g_{ji}^{\text{smc}}(\bar{q})|\boldsymbol{\sigma}_{\alpha\beta}\cdot(\boldsymbol{q}\times\boldsymbol{\epsilon}_q)|^2\sum_{i\nu_\ell}\mathscr{G}_j^{(0)}(i\omega_n - i\nu_\ell)\mathscr{D}_q^{(0)}(i\nu_\ell), \tag{S.8d}$$

where $g_{ij}^{\text{conv(smc)}}(q) = g^{\text{conv(smc)}}\sum_l \psi_i^*(\boldsymbol{R}_l)\psi_j(\boldsymbol{R}_l) = \left[g_{ji}^{\text{conv(smc)}}(\bar{q})\right]^*$. $\mathscr{G}_i^{(0)}(i\omega_n) = 1/(i\omega_n - \varepsilon_i)$ is the electron- and $\mathscr{D}_q^{(0)}(i\nu_\ell) = 2\hbar\omega_q/[(i\nu_\ell)^2 - (\hbar\omega_q)^2]$ is the phonon-propagators; $\omega_n$ and $\nu_\ell$ are fermionic and bosonic Matsubara frequencies, respectively. We further defined the extra phonon-propagators as $\mathscr{D}_q^{(\pm)}(i\nu_\ell) = \pm 2i\nu_\ell/[(i\nu_\ell)^2 - (\hbar\omega_q)^2]$. The hopping rate $\tau_{i\alpha}^{-1} = -\frac{2}{\hbar}\text{Im}\,\Sigma_{\alpha\alpha}(i,i;i\omega_n \to \varepsilon_i + i0)$ for each diagram is then obtained as

$$\frac{1}{\tau_{i\alpha}^{(a)}} = \frac{2\pi}{\hbar}\sum_{j,\beta}\sum_q |g_{ij}^{\text{conv}}(q)|^2\delta_{\alpha\beta}\Big[\{1 - f(\varepsilon_j) + n(\hbar\omega_q)\}\delta(\varepsilon_i - \varepsilon_j - \hbar\omega_q) + \{f(\varepsilon_j) + n(\hbar\omega_q)\}\delta(\varepsilon_i - \varepsilon_j + \hbar\omega_q)\Big], \tag{S.9a}$$

$$\frac{1}{\tau_{i\alpha}^{(b)}} = \frac{2\pi}{\hbar}\sum_{j,\beta}\sum_q g_{ij}^{\text{smc}}(q)g_{ij}^{\text{conv}*}(q)\boldsymbol{\sigma}_{\alpha\beta}\cdot(\boldsymbol{q}\times\boldsymbol{\epsilon}_q)\delta_{\alpha\beta}\Big[\{1 - f(\varepsilon_j) + n(\hbar\omega_q)\}\delta(\varepsilon_i - \varepsilon_j - \hbar\omega_q) - \{f(\varepsilon_j) + n(\hbar\omega_q)\}\delta(\varepsilon_i - \varepsilon_j + \hbar\omega_q)\Big], \tag{S.9b}$$

$$\frac{1}{\tau_{i\alpha}^{(c)}} = \frac{2\pi}{\hbar}\sum_{j,\beta}\sum_q g_{ij}^{\text{conv}}(q)g_{ij}^{\text{smc}*}(q)\boldsymbol{\sigma}_{\alpha\beta}\cdot(\boldsymbol{q}\times\boldsymbol{\epsilon}_q)\delta_{\alpha\beta}\Big[\{1 - f(\varepsilon_j) + n(\hbar\omega_q)\}\delta(\varepsilon_i - \varepsilon_j - \hbar\omega_q) - \{f(\varepsilon_j) + n(\hbar\omega_q)\}\delta(\varepsilon_i - \varepsilon_j + \hbar\omega_q)\Big], \tag{S.9c}$$

$$\frac{1}{\tau_{i\alpha}^{(d)}} = \frac{2\pi}{\hbar}\sum_{j,\beta}\sum_q |g_{ij}^{\text{smc}}(q)|^2|\boldsymbol{\sigma}_{\alpha\beta}\cdot(\boldsymbol{q}\times\boldsymbol{\epsilon}_q)|^2\Big[\{1 - f(\varepsilon_j) + n(\hbar\omega_q)\}\delta(\varepsilon_i - \varepsilon_j - \hbar\omega_q) + \{f(\varepsilon_j) + n(\hbar\omega_q)\}\delta(\varepsilon_i - \varepsilon_j + \hbar\omega_q)\Big], \tag{S.9d}$$

where we have used the following relations:

$$-\frac{1}{\beta}\sum_{i\nu_\ell}\mathscr{G}_l^{(0)}(i\omega_n - i\nu_\ell)\mathscr{D}_q^{(0)}(i\nu_\ell) = \frac{1 - f(\varepsilon_l) + n(\hbar\omega_q)}{i\omega_n - \varepsilon_l - \hbar\omega_q} + \frac{f(\varepsilon_l) + n(\hbar\omega_q)}{i\omega_n - \varepsilon_l + \hbar\omega_q} \quad \text{(S.10a)}$$

$$-\frac{1}{\beta}\sum_{i\nu_\ell}\mathscr{G}_l^{(0)}(i\omega_n - i\nu_\ell)\mathscr{D}_q^{(+)}(i\nu_\ell) = \frac{1 - f(\varepsilon_l) + n(\hbar\omega_q)}{i\omega_n - \varepsilon_l - \hbar\omega_q} - \frac{f(\varepsilon_l) + n(\hbar\omega_q)}{i\omega_n - \varepsilon_l + \hbar\omega_q}. \quad \text{(S.10b)}$$

By converting the sum $\sum_q = \int_0^\infty d\varepsilon\,\nu(\varepsilon)$ and using $|g_{ij}^{\text{conv}}(q)|^2 \approx g^2 e^{-2|\boldsymbol{R}_i - \boldsymbol{R}_j|/\xi}$ in Eq. (S.9a), we reproduce the conventional variable-range hopping rate Eq. (6) in the main text as

$$\frac{1}{\tau_{\text{VRH}}} \simeq g^2 \sum_R e^{-2R/\xi} e^{-\Delta_R/2k_B T} \nu(\Delta_R), \quad \text{(S.11)}$$

where $R = |\boldsymbol{R}_i - \boldsymbol{R}_j|$ and we have assumed $k_B T \ll \varepsilon_i, \varepsilon_j, \Delta_R = |\varepsilon_i - \varepsilon_j|$.

## RANDOM SPIN RESISTOR NETWORK

### I. Rate Equation

The rate equation for an electron with a state $(i, \alpha)$ is given by

$$\frac{df_{i\alpha}}{dt} = -\left(\sum_j \sum_\beta \Gamma_{(i\alpha)\to(j\beta)}\right) f_{i\alpha} + \sum_j \sum_\beta \Gamma_{(j\beta)\to(i\alpha)} f_{j\beta} = \sum_j \sum_\beta \left(-\Gamma_{(i\alpha)\to(j\beta)} f_{i\alpha} + \Gamma_{(j\beta)\to(i\alpha)} f_{j\beta}\right), \quad \text{(S.12)}$$

where $f_{i\alpha}$ is the probability for finding an electron in a state $(i, \alpha)$ and $\Gamma_{(i\alpha)\to(j\beta)}$ is the transition rate for a jump $(i, \alpha) \to (j, \beta)$, which can be calculated, e.g., from Fermi's golden rule.

The time average of the above equation reads

$$\left\langle \frac{df_{i\alpha}}{dt} \right\rangle_t = \sum_j \sum_\beta \left\langle -\Gamma_{(i\alpha)\to(j\beta)} f_{i\alpha} + \Gamma_{(j\beta)\to(i\alpha)} f_{j\beta} \right\rangle_t. \quad \text{(S.13)}$$

Then, the time-averaged net charge flow from site $j$ with spin $\beta$ to site $i$ with spin $\alpha$ is given by

$$I_{(j\beta)\to(i\alpha)} := (-e)\left(-\langle \Gamma_{(i\alpha)\to(j\beta)} f_{i\alpha} \rangle_t + \langle \Gamma_{(j\beta)\to(i\alpha)} f_{j\beta} \rangle_t\right). \quad \text{(S.14)}$$

In the absence of an electric field $\boldsymbol{E}$ (in equilibrium) which we denote by the subscript "0", electron transfer is random and there is a detailed balance in its time average; therefore, no net charge/spin current survives. Namely, $\langle \Gamma_{(i\alpha)\to(j\beta)} f_{i\alpha} \rangle_t^0$ must be symmetric with respect to the states $(i, \alpha)$ and $(j, \beta)$:

$$\langle \Gamma_{(i\alpha)\to(j\beta)} f_{i\alpha} \rangle_t^0 = \langle \Gamma_{(j\beta)\to(i\alpha)} f_{j\beta} \rangle_t^0. \quad \text{(S.15)}$$

By defining the intrinsic transition rate $\gamma_{(i\alpha)\to(j\beta)}$ as $\Gamma_{(i\alpha)\to(j\beta)} =: \gamma_{(i\alpha)\to(j\beta)}(1 - f_{j\beta})$, the above condition can be rewritten as

$$\langle \gamma_{(i\alpha)\to(j\beta)}(1 - f_{j\beta}) f_{i\alpha} \rangle_t^0 = \langle \gamma_{(j\beta)\to(i\alpha)}(1 - f_{i\alpha}) f_{j\beta} \rangle_t^0. \quad \text{(S.16)}$$

Throughout this work, we shall neglect the electron-electron interactions and then, $\gamma_{(i\alpha)\to(j\beta)}$ is independent of the distribution and may be removed from the brackets. Furthermore, in thermal equilibrium, the distribution for different states are statistically independent, so that $\langle f_{i\alpha} f_{j\beta} \rangle_t^0 = \langle f_{i\alpha} \rangle_t^0 \langle f_{j\beta} \rangle_t^0$ and $\langle f_{i\alpha} \rangle_t^0 =: f_{i\alpha}^0 = f(\varepsilon_i) = [e^{\beta\varepsilon_i} + 1]^{-1}$. Here, the energy $\varepsilon_i$ is measured from the Fermi level. Finally, from the detailed balance Eq. (S.15), we obtain the condition for the intrinsic transition rate:

$$\gamma_{(i\alpha)\to(j\beta)}^0 = \gamma_{(j\beta)\to(i\alpha)}^0 e^{\beta(\varepsilon_i - \varepsilon_j)}, \quad \text{(S.17)}$$

which is also satisfied for the transition rates calculated microscopically in Eqs. (S.9).

An external electric field modulates both the electron distribution and the intrinsic transition rate as follows:

$$\langle f_{i\alpha}\rangle_t = f_{i\alpha}^0 + \langle \delta f_{i\alpha}\rangle_t := [e^{\beta(\varepsilon_{i\alpha}-\delta\mu_i^\alpha)}+1]^{-1}, \qquad \gamma_{(i\alpha)\to(j\beta)} = \gamma_{(i\alpha)\to(j\beta)}^0 + \delta\gamma_{(i\alpha)\to(j\beta)}. \tag{S.18}$$

In the linear response regime, the time-averaged net charge flow can be approximated as,

$$\begin{aligned}
I_{(j\beta)\to(i\alpha)}/e &= \langle f_{i\alpha}\rangle_t(1-\langle f_{j\beta}\rangle_t)\gamma_{(i\alpha)\to(j\beta)} - \langle f_{j\beta}\rangle_t(1-\langle f_{i\alpha}\rangle_t)\gamma_{(j\beta)\to(i\alpha)} \\
&= (f_{i\alpha}^0 + \langle\delta f_{i\alpha}\rangle_t)(1-f_{j\beta}^0-\langle\delta f_{j\beta}\rangle_t)(\gamma_{(i\alpha)\to(j\beta)}^0+\delta\gamma_{(i\alpha)\to(j\beta)}) \\
&\quad - (f_{j\beta}^0 + \langle\delta f_{j\beta}\rangle_t)(1-f_{i\alpha}^0-\langle\delta f_{i\alpha}\rangle_t)(\gamma_{(j\beta)\to(i\alpha)}^0+\delta\gamma_{(j\beta)\to(i\alpha)}) \\
&= \Gamma_{(j\beta)\to(i\alpha)}^0 f_{j\beta}^0 \left[\frac{\delta\gamma_{(i\alpha)\to(j\beta)}}{\gamma_{(i\alpha)\to(j\beta)}^0} - \frac{\delta\gamma_{(j\beta)\to(i\alpha)}}{\gamma_{(j\beta)\to(i\alpha)}^0} + \frac{\langle\delta f_{i\alpha}\rangle_t}{f_{i\alpha}^0(1-f_{i\alpha}^0)} - \frac{\langle\delta f_{j\beta}\rangle_t}{f_{j\beta}^0(1-f_{j\beta}^0)}\right] + O(\boldsymbol{E}^2) \\
&\simeq \frac{1}{k_\text{B}T}\Gamma_{(j\beta)\to(i\alpha)}^0 f_{j\beta}^0[e\boldsymbol{E}\cdot(\boldsymbol{R}_i-\boldsymbol{R}_j)+\delta\mu_i^\alpha-\delta\mu_j^\beta],
\end{aligned} \tag{S.19}$$

where we have used the detailed balance condition $\Gamma_{(i\alpha)\to(j\beta)}^0 f_{i\alpha}^0 = \Gamma_{(j\beta)\to(i\alpha)}^0 f_{j\beta}^0$ and Eq. (S.17) to evaluate the terms involving $\delta\gamma_{(i\alpha)\to(j\beta)}$. Defining the spin-dependent electrochemical potential at each site as $-eV_i^\alpha := e\boldsymbol{E}\cdot\boldsymbol{R}_i+\delta\mu_i^\alpha$, the final form of the net charge flow from $(j,\beta)$ to $(i,\alpha)$ in the linear response to an external electric field is given by,

$$I_{(j\beta)\to(i\alpha)} = \frac{e^2}{k_\text{B}T}\Gamma_{(j\beta)\to(i\alpha)}^0 f_{j\beta}^0(V_j^\beta-V_i^\alpha) = G_{ji}^{\beta\alpha}(V_j^\beta-V_i^\alpha). \tag{S.20}$$

Here, we have defined the conductance as

$$G_{ji}^{\beta\alpha} := \frac{e^2}{k_\text{B}T}\Gamma_{(j\beta)\to(i\alpha)}^0 f_{j\beta}^0, \tag{S.21}$$

which is, by definition, also symmetric with respect to the states: $G_{ij}^{\alpha\beta} = G_{ji}^{\beta\alpha}$.

## II. Reduction to the Percolation Theory

We next must predict a more detailed form for $\gamma_{(i\alpha)\to(j\beta)}$. Because we are considering a tunneling process, we know that the dominant dependence of $\gamma_{(i\alpha)\to(j\beta)}$ on $R_{ij} := |\boldsymbol{R}_i-\boldsymbol{R}_j|$ must be exponential,

$$\gamma_{(i\alpha)\to(j\beta)} \propto e^{-2R_{ij}/\xi}, \tag{S.22}$$

where $\xi$ is the localization length.

The energy dependence of $\gamma_{(i\alpha)\to(j\beta)}$ is less obvious than the $R$-dependence; and, in fact, a number of different kinds of behavior seem possible. The simplest situation occurs when $k_\text{B}T$ is small compared to $|\varepsilon_i-\varepsilon_j|$, and the energy difference $|\varepsilon_i-\varepsilon_j|$ is of the order of the Debye energy or smaller. It is then a good approximation to write

$$\gamma_{(i\alpha)\to(j\beta)}^0 = \gamma_{\alpha\beta}^0 \times \begin{cases} e^{-2R_{ij}/\xi}e^{-(\varepsilon_j-\varepsilon_i)/k_\text{B}T} & \varepsilon_j > \varepsilon_i \\ e^{-2R_{ij}/\xi} & \varepsilon_j < \varepsilon_i \end{cases}, \tag{S.23}$$

where $\gamma_{\alpha\beta}^0$ is some constant which depends on the electron-phonon coupling strength, the phonon density of states, and other properties of the material. Eq. (S.23) satisfies the detailed balance condition Eq. (S.17).

Combining Eqs. (S.23) and assuming $k_\text{B}T$ small compared to all energies, we find that the value of $\Gamma_{ij}$ in thermal equilibrium can be written in the relatively simple form:

$$\Gamma_{(i\alpha)\to(j\beta)}^0 f_{i\alpha}^0 = \gamma_{\alpha\beta}^0 e^{-2R_{ij}/\xi}\frac{e^{(\varepsilon_i+\varepsilon_j-|\varepsilon_i-\varepsilon_j|)/2k_\text{B}T}}{[1+e^{\varepsilon_i/k_\text{B}T}][1+e^{\varepsilon_j/k_\text{B}T}]} \tag{S.24}$$

$$\simeq \gamma_{\alpha\beta}^0\exp\left[-\frac{2R_{ij}}{\xi}-\frac{|\varepsilon_i|+|\varepsilon_j|+|\varepsilon_i-\varepsilon_j|}{2k_\text{B}T}\right]. \tag{S.25}$$

The most important parameters in this expression are the random variables $\varepsilon_i$, $\varepsilon_j$, and $R_{ij}$ in the exponential, which lead to a wide spread in its distribution of magnitude.

The application of percolation concepts to the random resistor network is straightforward. We here define the critical percolation conductance $G_c$ for a random network, where the values of the individual conductances vary over many orders of magnitude, as the largest value of the conductance such that the subset of resistors with $G_{ij} > G_c$ still contains a connected network which spans the entire system.

Then, the relation to the bond percolation problem is established by the following assumption:

$$i \text{ and } j \text{ is } \begin{cases} \text{disconnected} & (G_{ij} < G_c) \\ \text{connected with } G_{ij} & (G_{ij} > G_c) \end{cases}, \iff i \text{ and } j \text{ is } \begin{cases} \text{disconnected} & (\eta_{ij} < \eta_c) \\ \text{connected with } G_{ij} & (\eta_{ij} > \eta_c) \end{cases}, \quad (S.26)$$

where we have introduced the exponential factors $G_{ij} = G_0 e^{\eta_{ij}}$ and $G_c = G_0 e^{\eta_c}$. Then, the solution exists only when a continuous path cross the network from one end to another.

### III. Generalized Kirchhoff's Law

For given network conditions, the current circuit problem can be solved with the Kirchhoff's law. The generalized Kirchhoff's law including spin components is given by,

$$\sum_{j \in Z(i)} \sum_\beta G_{ij}^{\alpha\beta}(V_j^\beta - V_i^\alpha) + (I_i^\alpha)^{\text{source}} = 0, \quad (S.27a)$$

or equivalently,

$$\sum_{j \in Z(i)} \begin{bmatrix} G_{ij}^{\uparrow\uparrow} & G_{ij}^{\uparrow\downarrow} \\ G_{ij}^{\downarrow\uparrow} & G_{ij}^{\downarrow\downarrow} \end{bmatrix} \begin{bmatrix} V_j^\uparrow \\ V_j^\downarrow \end{bmatrix} + \begin{bmatrix} -\sum_{j \in Z(i)}(G_{ij}^{\uparrow\uparrow} + G_{ij}^{\uparrow\downarrow}) & 0 \\ 0 & -\sum_{j \in Z(i)}(G_{ij}^{\downarrow\uparrow} + G_{ij}^{\downarrow\downarrow}) \end{bmatrix} \begin{bmatrix} V_i^\uparrow \\ V_i^\downarrow \end{bmatrix} + \begin{bmatrix} I_i^\uparrow \\ I_i^\downarrow \end{bmatrix}^{\text{source}} = 0, \quad (S.27b)$$

where the sum runs over the set of all nodes $Z(i)$ connected to the node $i$. Due to the spin-preserving nature of the nodes, the incoming current is equal to the outgoing current at each node for both spin components. $(I_i^\alpha)^{\text{source}}$ represent external source currents supplied by the battery.

The above Kirchhoff's law Eq. (S.27) for each spin component of the charge currents at a node $i$. For the purpose of corresponding to the experimental conditions, it would be more prospective to transform it into the charge-spin basis. To this end, we multiply the above matrix equation by the transformation matrix $\begin{bmatrix} 1 & 1 \\ 1 & -1 \end{bmatrix}$ from the left and obtain the following equation:

$$\sum_{j \in Z(i)} \mathbf{G}_{ij} \mathbf{V}_j + \mathbf{G}_{ii} \mathbf{V}_i + \mathbf{I}_i^{\text{source}} = 0, \quad (S.28a)$$

where we have introduced the conductance matrices in the charge-spin basis as

$$\begin{aligned}
\mathbf{G}_{ij} &= \begin{bmatrix} 1 & 1 \\ 1 & -1 \end{bmatrix} \begin{bmatrix} G_{ij}^{\uparrow\uparrow} & G_{ij}^{\uparrow\downarrow} \\ G_{ij}^{\downarrow\uparrow} & G_{ij}^{\downarrow\downarrow} \end{bmatrix} \begin{bmatrix} 1 & 1 \\ 1 & -1 \end{bmatrix} \\
&= \begin{bmatrix} G_{ij}^{\uparrow\uparrow} + G_{ij}^{\uparrow\downarrow} + G_{ij}^{\downarrow\uparrow} + G_{ij}^{\downarrow\downarrow} & G_{ij}^{\uparrow\uparrow} + G_{ij}^{\downarrow\uparrow} - G_{ij}^{\uparrow\downarrow} - G_{ij}^{\downarrow\downarrow} \\ G_{ij}^{\uparrow\uparrow} + G_{ij}^{\uparrow\downarrow} - G_{ij}^{\downarrow\uparrow} - G_{ij}^{\downarrow\downarrow} & G_{ij}^{\uparrow\uparrow} + G_{ij}^{\downarrow\downarrow} - G_{ij}^{\uparrow\downarrow} - G_{ij}^{\downarrow\uparrow} \end{bmatrix} \\
&=: \begin{bmatrix} G_{ij}^{\text{cc}} & G_{ij}^{\text{cs}} \\ G_{ij}^{\text{sc}} & G_{ij}^{\text{ss}} \end{bmatrix},
\end{aligned} \quad (S.28b)$$

$$\begin{aligned}
\mathbf{G}_{ii} &= -\sum_{j \in Z(i)} \begin{bmatrix} 1 & 1 \\ 1 & -1 \end{bmatrix} \begin{bmatrix} G_{ij}^{\uparrow\uparrow} + G_{ij}^{\uparrow\downarrow} & 0 \\ 0 & G_{ij}^{\downarrow\uparrow} + G_{ij}^{\downarrow\downarrow} \end{bmatrix} \begin{bmatrix} 1 & 1 \\ 1 & -1 \end{bmatrix} \\
&= -\sum_{j \in Z(i)} \begin{bmatrix} G_{ij}^{\uparrow\uparrow} + G_{ij}^{\uparrow\downarrow} + G_{ij}^{\downarrow\uparrow} + G_{ij}^{\downarrow\downarrow} & G_{ij}^{\uparrow\uparrow} + G_{ij}^{\uparrow\downarrow} - G_{ij}^{\downarrow\uparrow} - G_{ij}^{\downarrow\downarrow} \\ G_{ij}^{\uparrow\uparrow} + G_{ij}^{\uparrow\downarrow} - G_{ij}^{\downarrow\uparrow} - G_{ij}^{\downarrow\downarrow} & G_{ij}^{\uparrow\uparrow} + G_{ij}^{\uparrow\downarrow} + G_{ij}^{\downarrow\uparrow} + G_{ij}^{\downarrow\downarrow} \end{bmatrix} \\
&= -\sum_{j \in Z(i)} \begin{bmatrix} G_{ij}^{\text{cc}} & G_{ij}^{\text{sc}} \\ G_{ij}^{\text{sc}} & G_{ij}^{\text{cc}} \end{bmatrix} \\
&=: \begin{bmatrix} G_{ii}^{\text{cc}} & G_{ii}^{\text{cs}} \\ G_{ii}^{\text{sc}} & G_{ii}^{\text{ss}} \end{bmatrix},
\end{aligned} \quad (S.28c)$$

and the charge/spin voltages and currents as

$$\mathbf{V}_i = \frac{1}{2}\begin{bmatrix} 1 & 1 \\ 1 & -1 \end{bmatrix}\begin{bmatrix} V_i^\uparrow \\ V_i^\downarrow \end{bmatrix} =: \begin{bmatrix} V_i^c \\ V_i^s \end{bmatrix}, \qquad \mathbf{I}_i = \begin{bmatrix} 1 & 1 \\ 1 & -1 \end{bmatrix}\begin{bmatrix} I_i^\uparrow \\ I_i^\downarrow \end{bmatrix} =: \begin{bmatrix} I_i^c \\ I_i^s \end{bmatrix}. \tag{S.28d}$$

Here, we have used the fact that $\frac{1}{2}\begin{bmatrix} 1 & 1 \\ 1 & -1 \end{bmatrix}^2 = \begin{bmatrix} 1 & 0 \\ 0 & 1 \end{bmatrix}$. As can be seen in the above transformation, the definitions of charge/spin currents and voltages naturally emerge:

$$V_i^{c/s} := \frac{V_i^\uparrow \pm V_i^\downarrow}{2}, \qquad I_i^{c/s} := I_i^\uparrow \pm I_i^\downarrow, \tag{S.28e}$$

and therefore, charge and spin chemical potentials (or spin accumulation) also appears, $\mu_i^{c/s} := (\mu_i^\uparrow \pm \mu_i^\downarrow)/2$.

By definition, the components of the conductance matrix obey the following relations [S.2]:

$$G_{ij}^{cc} = G_{ji}^{cc}, \quad G_{ij}^{ss} = G_{ji}^{ss}, \quad G_{ij}^{cs} = G_{ji}^{sc}, \qquad (j \neq i) \tag{S.29a}$$

$$G_{ii}^{cc} = G_{ii}^{ss}, \quad G_{ii}^{sc} = G_{ii}^{cs}, \tag{S.29b}$$

$$G_{ij}^{cs} = G_{ij}^{sc} - 2(G_{ij}^{\uparrow\downarrow} - G_{ij}^{\downarrow\uparrow}), \tag{S.29c}$$

$$G_{ij}^{ss} = G_{ij}^{cc} - 2(G_{ij}^{\uparrow\downarrow} + G_{ij}^{\downarrow\uparrow}). \tag{S.29d}$$

These constrains reduce the number of independent components of $\mathbf{G}_{ij}$.

We can write the matrix equations similar to Eq. (S.28a) for all nodal potentials ranging from $\mathbf{V}_1$ to $\mathbf{V}_N$ and cast those equations into the following matrix form:

$$\begin{bmatrix} \mathbf{L}_{11} & \mathbf{L}_{12} & \cdots & \mathbf{L}_{1N} \\ \mathbf{L}_{21} & \mathbf{L}_{22} & \cdots & \mathbf{L}_{2N} \\ \vdots & \vdots & \ddots & \vdots \\ \mathbf{L}_{N1} & \mathbf{L}_{N2} & \cdots & \mathbf{L}_{NN} \end{bmatrix}\begin{bmatrix} \mathbf{V}_1 \\ \mathbf{V}_2 \\ \vdots \\ \mathbf{V}_N \end{bmatrix} + \begin{bmatrix} \mathbf{I}_1 \\ \mathbf{I}_2 \\ \vdots \\ \mathbf{I}_N \end{bmatrix}^{\text{source}} = 0, \tag{S.30}$$

where $\mathbf{L}_{ij}$ are $2\times 2$ matrices and defined as

$$\mathbf{L}_{ij} = \begin{cases} \mathbf{G}_{ii} & i = j \\ \mathbf{G}_{ij} & i \neq j, \quad j \in Z(i) \\ 0 & \text{otherwise} \end{cases}. \tag{S.31}$$

From the conditions Eqs. (S.29a) and (S.29b), the weighted Laplacian matrix,

$$\mathbb{L} := \begin{bmatrix} \mathbf{L}_{11} & \mathbf{L}_{12} & \cdots & \mathbf{L}_{1N} \\ \mathbf{L}_{21} & \mathbf{L}_{22} & \cdots & \mathbf{L}_{2N} \\ \vdots & \vdots & \ddots & \vdots \\ \mathbf{L}_{N1} & \mathbf{L}_{N2} & \cdots & \mathbf{L}_{NN} \end{bmatrix}, \tag{S.32}$$

is symmetric, $\mathbb{L}^T = \mathbb{L}$.

## IV. Sum Rules

In this section, we derive the sum rules which are physically justified by starting from the generalized Kirchhoff's law:

$$\sum_j \begin{bmatrix} L_{ij}^{cc} & L_{ij}^{cs} \\ L_{ij}^{sc} & L_{ij}^{ss} \end{bmatrix}\begin{bmatrix} V_j^c \\ V_j^s \end{bmatrix} + \begin{bmatrix} I_i^c \\ I_i^s \end{bmatrix}^{\text{source}} = 0. \tag{S.33}$$

Regardless of how charge currents are generated by a charge voltage through $L^{cc}$ or by a spin voltage $L^{cs}$, charge conservation law requires them to add up to zero at steady state:

$$\sum_j \left(\sum_i L_{ij}^{cc}\right) V_j^c + \sum_j \left(\sum_i L_{ij}^{cs}\right) V_j^s = 0. \tag{S.34}$$

This requires two universal sum rules to hold:

- Sum Rule #1

$$\sum_i L_{ij}^{cc} = \sum_{i \in Z(j)} G_{ij}^{cc} + G_{jj}^{cc}$$
$$= \sum_{i \in Z(j)} (G_{ij}^{\uparrow\uparrow} + G_{ij}^{\uparrow\downarrow} + G_{ij}^{\downarrow\uparrow} + G_{ij}^{\downarrow\downarrow}) - \sum_{i \in Z(j)} (G_{ji}^{\uparrow\uparrow} + G_{ji}^{\uparrow\downarrow} + G_{ji}^{\downarrow\uparrow} + G_{ji}^{\downarrow\downarrow})$$
$$= 0, \qquad (G_{ij}^{\alpha\beta} = G_{ji}^{\beta\alpha}) \tag{S.35}$$

- Sum Rule #2

$$\sum_i L_{ij}^{cs} = \sum_{i \in Z(j)} G_{ij}^{cs} + G_{jj}^{cs}$$
$$= \sum_{i \in Z(j)} (G_{ij}^{\uparrow\uparrow} + G_{ij}^{\downarrow\uparrow} - G_{ij}^{\uparrow\downarrow} - G_{ij}^{\downarrow\downarrow}) - \sum_{i \in Z(j)} (G_{ji}^{\uparrow\uparrow} + G_{ji}^{\uparrow\downarrow} - G_{ji}^{\downarrow\uparrow} - G_{ji}^{\downarrow\downarrow})$$
$$= 0. \tag{S.36}$$

In equilibrium, there are no spin accumulation: $\mu_j^s = 0$. Therefore,

- Sum Rule #3

$$\sum_j L_{ij}^{cc} = 0, \tag{S.37}$$

guarantees that the absence of any charge current at each node in equilibrium with $V_j^c = $ const. and $V_j^s = 0$.

Similarly,

- Sum Rule #4

$$\sum_j L_{ij}^{sc} = \sum_{j \in Z(j)} G_{ij}^{sc} + G_{ii}^{sc}$$
$$= \sum_{j \in Z(i)} (G_{ij}^{\uparrow\uparrow} + G_{ij}^{\uparrow\downarrow} - G_{ij}^{\downarrow\uparrow} - G_{ij}^{\downarrow\downarrow}) - \sum_{j \in Z(i)} (G_{ij}^{\uparrow\uparrow} + G_{ij}^{\uparrow\downarrow} - G_{ij}^{\downarrow\uparrow} - G_{ij}^{\downarrow\downarrow})$$
$$= 0, \tag{S.38}$$

also guarantees the absence of any spin current at each node in equilibrium [S.3].

We also note that there are no general sum rules for the spin-to-spin conductance $L_{ij}^{ss}$. This is closely related to the fact that the spin current is not conserved through hopping.

Finally, from the sum rules #3 and #4, the generalized Kirhhoff's law Eq. (S.33) does not change its form for any constant shift of the charge voltage $V_j^c \to V_j^c + V_0$:

$$\sum_j L_{ij}^{cc}(V_j^c + V_0) + \sum_j L_{ij}^{cs} V_j^s + (I_i^c)^{\text{source}} = \sum_j L_{ij}^{cc} V_j^c + \sum_j L_{ij}^{cs} V_j^s + (I_i^c)^{\text{source}} = 0,$$
$$\sum_j L_{ij}^{sc}(V_j^c + V_0) + \sum_j L_{ij}^{ss} V_j^s + (I_i^s)^{\text{source}} = \sum_j L_{ij}^{sc} V_j^c + \sum_j L_{ij}^{ss} V_j^s + (I_i^s)^{\text{source}} = 0.$$

Therefore, we can chose $V_N^c = 0$ by grounding one end of the circuit.

### V. Boundary Conditions and Equivalent Conductance

The next step is to compute the physical quantities such as the charge conductance $G^c := I/V$. Our starting point is the weighted Laplacian matrix $\mathbb{L}$ associated with a given network, whose entries are the conductances $\mathbf{L}_{ij}$ connecting pairs of nodes. We assume that two nodes at the ends of the system are connected to the battery, which

fixes the charge voltage difference $V$ between these two nodes. Then, the battery terminals read $V_1^c = V$ and $V_N^c = 0$ and we can compute the observables from $\mathbb{L}$ and its relatives.

To this end, the boundary conditions with the external environment become important. First, the constraints for intermediate nodes $i = 2, 3, \cdots, N-1$, which are not connected to the environment, are given by,

$$(I_i^c)^{\text{source}} = 0, \qquad (I_i^s)^{\text{source}} = 0. \tag{S.39a}$$

Second, the boundary conditions for the charge components stemming from an external battery are given by,

$$I_1^c = I, \qquad V_1^c = V, \tag{S.39b}$$
$$I_N^c = -I, \quad V_N^c = 0. \tag{S.39c}$$

Finally, we further take the following boundary conditions for the spin components:

$$(I_1^s)^{\text{source}} = (I_N^s)^{\text{source}} = 0. \tag{S.39d}$$

If one wants to correspond to more typical experimental conditions that DNA is directly contacted with a nonmagnetic electrode such as Au on one side and a ferromagnetic electrode such as Ni on the other side, for example, one should set the boundary condition for the spin injection as $I_1^s = \alpha_{\text{Ni}} I$. Here, $\alpha_{\text{Ni}} \simeq 0.23$ is the spin polarization ratio of Ni. In this way, we can represent the injection of a spin-polarized current from a ferromagnetic electrode without adding extra degrees of freedom.

By combining Eq. (S.30) with Eqs. (S.39), we obtain the following matrix equations:

$$\begin{bmatrix} \begin{bmatrix} L_{11}^{cc} & L_{11}^{cs} \\ L_{11}^{sc} & L_{11}^{ss} \end{bmatrix} & \begin{bmatrix} L_{12}^{cc} & L_{12}^{cs} \\ L_{12}^{sc} & L_{12}^{ss} \end{bmatrix} & \cdots & \begin{bmatrix} L_{1N}^{cc} & L_{1N}^{cs} \\ L_{1N}^{sc} & L_{1N}^{ss} \end{bmatrix} \\ \begin{bmatrix} L_{21}^{cc} & L_{21}^{cs} \\ L_{21}^{sc} & L_{21}^{ss} \end{bmatrix} & \begin{bmatrix} L_{22}^{cc} & L_{22}^{cs} \\ L_{22}^{sc} & L_{22}^{ss} \end{bmatrix} & \cdots & \begin{bmatrix} L_{2N}^{cc} & L_{2N}^{cs} \\ L_{2N}^{sc} & L_{2N}^{ss} \end{bmatrix} \\ \vdots & \vdots & \ddots & \vdots \\ \begin{bmatrix} L_{N1}^{cc} & L_{N1}^{cs} \\ L_{N1}^{sc} & L_{N1}^{ss} \end{bmatrix} & \begin{bmatrix} L_{N2}^{cc} & L_{N2}^{cs} \\ L_{N2}^{sc} & L_{N2}^{ss} \end{bmatrix} & \cdots & \begin{bmatrix} L_{NN}^{cc} & L_{NN}^{cs} \\ L_{NN}^{sc} & L_{NN}^{ss} \end{bmatrix} \end{bmatrix} \begin{bmatrix} \begin{bmatrix} V \\ V_1^s \end{bmatrix} \\ \begin{bmatrix} V_2^c \\ V_2^s \end{bmatrix} \\ \vdots \\ \begin{bmatrix} 0 \\ V_N^s \end{bmatrix} \end{bmatrix} + \begin{bmatrix} \begin{bmatrix} I \\ 0 \end{bmatrix} \\ \begin{bmatrix} 0 \\ 0 \end{bmatrix} \\ \vdots \\ \begin{bmatrix} -I \\ 0 \end{bmatrix} \end{bmatrix} = 0, \tag{S.40}$$

where the unknown quantities are $I, V_1^s, V_2^c, V_2^s, \cdots, V_{N-1}^c, V_{N-1}^s$, and $V_N^s$. Note that the sum rules #1 and #2 imply that the 2N equations in Eq. (S.40) are not independent. Hence, from now on, we will skip the $(2N-1)$-th equation related to $-I$ and focus only on the remaining $(2N-1)$ equations.

In order to solve for the unknown quantities, we rearrange Eq. (S.40) as follows. The first element of each row of $\mathbb{L}$ multiplies $V$. We carry this term to the right-hand side of each equation. In the first row, we also carry the unknown input charge current $I$ to the left-hand side. Now these equations take the form:

$$\begin{bmatrix} \begin{bmatrix} 1 & L_{11}^{cs} \\ 0 & L_{11}^{ss} \end{bmatrix} & \begin{bmatrix} L_{12}^{cc} & L_{12}^{cs} \\ L_{12}^{sc} & L_{12}^{ss} \end{bmatrix} & \cdots & \begin{bmatrix} L_{1N}^{cs} \\ L_{1N}^{ss} \end{bmatrix} \\ \begin{bmatrix} 0 & L_{21}^{cs} \\ 0 & L_{21}^{ss} \end{bmatrix} & \begin{bmatrix} L_{22}^{cc} & L_{22}^{cs} \\ L_{22}^{sc} & L_{22}^{ss} \end{bmatrix} & \cdots & \begin{bmatrix} L_{2N}^{cs} \\ L_{2N}^{ss} \end{bmatrix} \\ \vdots & \vdots & \ddots & \vdots \\ \begin{bmatrix} 0 & L_{N1}^{ss} \end{bmatrix} & \begin{bmatrix} L_{N2}^{sc} & L_{N2}^{ss} \end{bmatrix} & \cdots & L_{NN}^{ss} \end{bmatrix} \begin{bmatrix} \begin{bmatrix} I \\ V_1^s \end{bmatrix} \\ \begin{bmatrix} V_2^c \\ V_2^s \end{bmatrix} \\ \vdots \\ V_N^s \end{bmatrix} = -V \begin{bmatrix} \begin{bmatrix} L_{11}^{cc} \\ L_{11}^{sc} \end{bmatrix} \\ \begin{bmatrix} L_{21}^{cc} \\ L_{21}^{sc} \end{bmatrix} \\ \vdots \\ L_{N1}^{sc} \end{bmatrix}. \tag{S.41}$$

Applying Cramer's rule, we then obtain expressions for the equivalent charge conductance and the nodal spin voltages as

$$G_{\text{eq}}^c = -\frac{\det \mathbb{L}'}{\det \mathbb{L}''}, \qquad V_i^s = -V \frac{\det \mathbb{L}_i''}{\det \mathbb{L}''} \quad (1 \leq i \leq N-1), \qquad V_N^s = V \frac{\det \mathbb{L}_N''}{\det \mathbb{L}''}. \tag{S.42}$$

Here, we have defined a $(2N-1) \times (2N-1)$ sub-matrix of $\mathbb{L}$,

$$\mathbb{L}' = \begin{bmatrix} \begin{bmatrix} L_{11}^{\rm cc} & L_{11}^{\rm cs} \\ L_{11}^{\rm sc} & L_{11}^{\rm ss} \end{bmatrix} & \begin{bmatrix} L_{12}^{\rm cc} & L_{12}^{\rm cs} \\ L_{12}^{\rm sc} & L_{12}^{\rm ss} \end{bmatrix} & \cdots & \begin{bmatrix} L_{1N-1}^{\rm cc} & L_{1N-1}^{\rm cs} \\ L_{1N-1}^{\rm sc} & L_{1N-1}^{\rm ss} \end{bmatrix} & \begin{bmatrix} L_{1N}^{\rm cs} \\ L_{1N}^{\rm ss} \end{bmatrix} \\ \begin{bmatrix} L_{21}^{\rm cc} & L_{21}^{\rm cs} \\ L_{21}^{\rm sc} & L_{21}^{\rm ss} \end{bmatrix} & \begin{bmatrix} L_{22}^{\rm cc} & L_{22}^{\rm cs} \\ L_{22}^{\rm sc} & L_{22}^{\rm ss} \end{bmatrix} & \cdots & \begin{bmatrix} L_{2N-1}^{\rm cc} & L_{2N-1}^{\rm cs} \\ L_{2N-1}^{\rm sc} & L_{2N-1}^{\rm ss} \end{bmatrix} & \begin{bmatrix} L_{2N}^{\rm cs} \\ L_{2N}^{\rm ss} \end{bmatrix} \\ \vdots & \vdots & \ddots & \vdots & \vdots \\ \begin{bmatrix} L_{N1}^{\rm sc} & L_{N1}^{\rm ss} \end{bmatrix} & \begin{bmatrix} L_{N2}^{\rm sc} & L_{N2}^{\rm ss} \end{bmatrix} & \cdots & \begin{bmatrix} L_{NN-1}^{\rm sc} & L_{NN-1}^{\rm ss} \end{bmatrix} & L_{NN}^{\rm ss} \end{bmatrix}, \quad \text{(S.43a)}$$

and its $(2N-2) \times (2N-2)$ sub-matrices,

$$\mathbb{L}'' = \begin{bmatrix} L_{11}^{\rm ss} & \begin{bmatrix} L_{12}^{\rm sc} & L_{12}^{\rm ss} \end{bmatrix} & \cdots & L_{1N}^{\rm ss} \\ \begin{bmatrix} L_{21}^{\rm cs} \\ L_{21}^{\rm ss} \end{bmatrix} & \begin{bmatrix} L_{22}^{\rm cc} & L_{22}^{\rm cs} \\ L_{22}^{\rm sc} & L_{22}^{\rm ss} \end{bmatrix} & \cdots & \begin{bmatrix} L_{2N}^{\rm cs} \\ L_{2N}^{\rm ss} \end{bmatrix} \\ \vdots & \vdots & \ddots & \vdots \\ L_{N1}^{\rm ss} & \begin{bmatrix} L_{N2}^{\rm sc} & L_{N2}^{\rm ss} \end{bmatrix} & \cdots & L_{NN}^{\rm ss} \end{bmatrix}, \quad \mathbb{L}''_i = \begin{bmatrix} \begin{bmatrix} L_{11}^{\rm sc} & L_{11}^{\rm ss} \end{bmatrix} & \cdots & L_{1i}^{\rm sc} & \begin{bmatrix} L_{1i+1}^{\rm sc} & L_{1i+1}^{\rm ss} \end{bmatrix} & \cdots & L_{1N}^{\rm ss} \\ \begin{bmatrix} L_{21}^{\rm cc} & L_{21}^{\rm cs} \\ L_{21}^{\rm sc} & L_{21}^{\rm ss} \end{bmatrix} & \cdots & \begin{bmatrix} L_{2i}^{\rm cc} \\ L_{2i}^{\rm sc} \end{bmatrix} & \begin{bmatrix} L_{2i+1}^{\rm cc} & L_{2i+1}^{\rm cs} \\ L_{2i+1}^{\rm sc} & L_{2i+1}^{\rm ss} \end{bmatrix} & \cdots & \begin{bmatrix} L_{2N}^{\rm cs} \\ L_{2N}^{\rm ss} \end{bmatrix} \\ \vdots & \ddots & \vdots & \vdots & \ddots & \vdots \\ \begin{bmatrix} L_{N1}^{\rm sc} & L_{N1}^{\rm ss} \end{bmatrix} & \cdots & L_{Ni}^{\rm sc} & \begin{bmatrix} L_{Ni+1}^{\rm sc} & L_{Ni+1}^{\rm ss} \end{bmatrix} & \cdots & L_{NN}^{\rm ss} \end{bmatrix}.$$

(S.43b)

### VI. Applications to Spin-Microrotation Coupling

By substituting $\Gamma^0_{(i\alpha)\to(j\beta)}$ for the spin-microrotation coupling into the conductances, we obtain the following form of them:

$$G_{ij}^{\uparrow\uparrow} = \frac{e^2}{k_{\rm B}T}\frac{2\pi}{\hbar}\sum_q \left[\left\{|g_{ij}^{\rm conv}(q)|^2 + |g_{ij}^{\rm smc}(q)|^2|(\boldsymbol{q}\times\boldsymbol{\epsilon}_q)_\parallel|^2\right\}F_{ij}^+(q) + 2\,{\rm Re}\left\{g_{ij}^{\rm smc}(q)g_{ij}^{{\rm conv}*}(q)(\boldsymbol{q}\times\boldsymbol{\epsilon}_q)_\parallel\right\}F_{ij}^-(q)\right],$$
(S.44a)

$$G_{ij}^{\downarrow\downarrow} = \frac{e^2}{k_{\rm B}T}\frac{2\pi}{\hbar}\sum_q \left[\left\{|g_{ij}^{\rm conv}(q)|^2 + |g_{ij}^{\rm smc}(q)|^2|(\boldsymbol{q}\times\boldsymbol{\epsilon}_q)_\parallel|^2\right\}F_{ij}^+(q) - 2\,{\rm Re}\left\{g_{ij}^{\rm smc}(q)g_{ij}^{{\rm conv}*}(q)(\boldsymbol{q}\times\boldsymbol{\epsilon}_q)_\parallel\right\}F_{ij}^-(q)\right],$$
(S.44b)

$$G_{ij}^{\uparrow\downarrow} = \frac{e^2}{k_{\rm B}T}\frac{2\pi}{\hbar}\sum_q |g_{ij}^{\rm smc}(q)|^2|(\boldsymbol{q}\times\boldsymbol{\epsilon}_q)_+|^2 F_{ij}^+(q) = G_{ji}^{\downarrow\uparrow}, \tag{S.44c}$$

$$G_{ij}^{\downarrow\uparrow} = \frac{e^2}{k_{\rm B}T}\frac{2\pi}{\hbar}\sum_q |g_{ij}^{\rm smc}(q)|^2|(\boldsymbol{q}\times\boldsymbol{\epsilon}_q)_-|^2 F_{ij}^+(q), \tag{S.44d}$$

where we have defined $(\boldsymbol{q}\times\boldsymbol{\epsilon}_q)_\pm := (\boldsymbol{q}\times\boldsymbol{\epsilon}_q)_x \pm i(\boldsymbol{q}\times\boldsymbol{\epsilon}_q)_y$ and used the following relation,

$$|(\boldsymbol{q}\times\boldsymbol{\epsilon}_q)_\pm|^2 = |(\boldsymbol{q}\times\boldsymbol{\epsilon}_q)_x|^2 + |(\boldsymbol{q}\times\boldsymbol{\epsilon}_q)_y|^2 \mp 2\,{\rm Re}[(\boldsymbol{q}\times\boldsymbol{\epsilon}_q)_x(\boldsymbol{q}\times\boldsymbol{\epsilon}_q^*)_y]. \tag{S.45}$$

Furthermore, we have defined an (anti)symmetric function $F_{ij}^\pm(q) = \pm F_{ji}^\pm(q)$ by using the Heaviside step function $\theta(x)$ as

$$F_{ij}^\pm(q) := [1-f(\varepsilon_j)]f(\varepsilon_i)\Big[\{1+n(\varepsilon_i-\varepsilon_j)\}\theta(\varepsilon_i-\varepsilon_j) \pm n(\varepsilon_j-\varepsilon_i)\theta(\varepsilon_j-\varepsilon_i)\Big]\delta(|\varepsilon_i-\varepsilon_j|-\hbar\omega_q). \tag{S.46}$$

Then, the resulting conductances in the charge-spin basis are given by,

$$G_{ij}^{\text{cc}} = G_{ij}^{\uparrow\uparrow} + G_{ij}^{\uparrow\downarrow} + G_{ij}^{\downarrow\uparrow} + G_{ij}^{\downarrow\downarrow}$$
$$= \frac{e^2}{k_\text{B} T} \frac{4\pi}{\hbar} \sum_q \left[ \left|g_{ij}^{\text{conv}}(q)\right|^2 + \left|g_{ij}^{\text{smc}}(q)\right|^2 \left\{\boldsymbol{q}^2 - (\boldsymbol{q}\cdot\boldsymbol{\epsilon}_q)(\boldsymbol{q}\cdot\boldsymbol{\epsilon}_q^*)\right\} \right] F_{ij}^+(q), \tag{S.47a}$$

$$G_{ij}^{\text{cs}} = G_{ij}^{\uparrow\uparrow} + G_{ij}^{\downarrow\uparrow} - G_{ij}^{\uparrow\downarrow} - G_{ij}^{\downarrow\downarrow}$$
$$= \frac{e^2}{k_\text{B} T} \frac{4\pi}{\hbar} \sum_q \left\{ 2\,\text{Re}\left[g_{ij}^{\text{smc}}(q) g_{ij}^{\text{conv}*}(q)(\boldsymbol{q}\times\boldsymbol{\epsilon}_q)_\|\right] F_{ij}^-(q) - \left|g_{ij}^{\text{smc}}(q)\right|^2 q_\| [\boldsymbol{q}\cdot\text{Im}(\boldsymbol{\epsilon}_q^*\times\boldsymbol{\epsilon}_q)] F_{ij}^+(q) \right\}, \tag{S.47b}$$

$$G_{ij}^{\text{sc}} = G_{ij}^{\uparrow\uparrow} + G_{ij}^{\uparrow\downarrow} - G_{ij}^{\downarrow\uparrow} - G_{ij}^{\downarrow\downarrow} = G_{ji}^{\text{cs}}$$
$$= \frac{e^2}{k_\text{B} T} \frac{4\pi}{\hbar} \sum_q \left\{ 2\,\text{Re}\left[g_{ij}^{\text{smc}}(q) g_{ij}^{\text{conv}*}(q)(\boldsymbol{q}\times\boldsymbol{\epsilon}_q)_\|\right] F_{ij}^-(q) + \left|g_{ij}^{\text{smc}}(q)\right|^2 q_\| [\boldsymbol{q}\cdot\text{Im}(\boldsymbol{\epsilon}_q^*\times\boldsymbol{\epsilon}_q)] F_{ij}^+(q) \right\}, \tag{S.47c}$$

$$G_{ij}^{\text{ss}} = G_{ij}^{\uparrow\uparrow} + G_{ij}^{\downarrow\downarrow} - G_{ij}^{\uparrow\downarrow} - G_{ij}^{\downarrow\uparrow}$$
$$= \frac{e^2}{k_\text{B} T} \frac{4\pi}{\hbar} \sum_q \left[ \left|g_{ij}^{\text{conv}}(q)\right|^2 + \left|g_{ij}^{\text{smc}}(q)\right|^2 \left(\left|(\boldsymbol{q}\times\boldsymbol{\epsilon}_q)_\|\right|^2 - \left|(\boldsymbol{q}\times\boldsymbol{\epsilon}_q)_x\right|^2 - \left|(\boldsymbol{q}\times\boldsymbol{\epsilon}_q)_y\right|^2 \right) \right] F_{ij}^+(q). \tag{S.47d}$$

Here, we have also used the relation: $|\boldsymbol{q}\times\boldsymbol{\epsilon}_q|^2 = \boldsymbol{q}^2(\boldsymbol{\epsilon}_q\cdot\boldsymbol{\epsilon}_q^*) - (\boldsymbol{q}\cdot\boldsymbol{\epsilon}_q)(\boldsymbol{q}\cdot\boldsymbol{\epsilon}_q^*) = \boldsymbol{q}^2 - (\boldsymbol{q}\cdot\boldsymbol{\epsilon}_q)(\boldsymbol{q}\cdot\boldsymbol{\epsilon}_q^*)$.

## TEMPERATURE FITTING

DNA is quite different from inorganic materials in that DNA chains are flexible and have strong structural fluctuations, which may crucially affect the transport properties. Strong structural fluctuations in DNA further localize electronic wave functions and results in a temperature-dependent localization length due to the thermal nature of these fluctuations. Then, according to Ref. [S.4], we introduce the temperature-dependent localization length $\xi^{-1}(T) =: \alpha(T) = \alpha_0 + \alpha_1 \tanh(T/T_\text{d})^2$ with $\alpha_0 = 0.18\,\text{Å}^{-1}$, $\alpha_1 = 0.70\,\text{Å}^{-1}$, and $T_\text{d} = 192\,\text{K}$ in the main text.

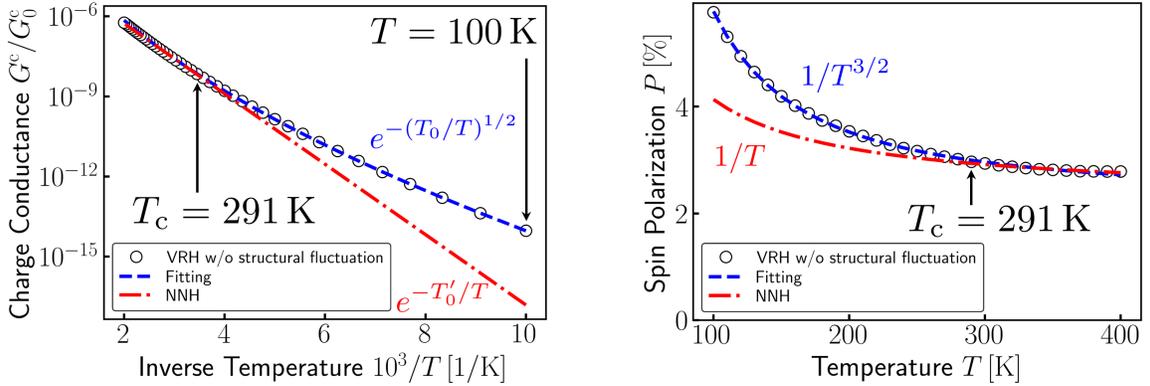

Supplementary Figure S2. Temperature dependences of (Left) the charge conductance and (Right) the spin polarization without structural fluctuations for a molecular length of 40 base-pairs.

We have performed numerical Monte Carlo simulations and investigated the temperature dependences of the observables. Fig. S2 shows the fitting results of the charge conductance $G^\text{c}$ and the spin polarization $P$ where electrons obey VRH without structural fluctuations, $\xi^{-1} = 0.88\,\text{Å}^{-1}$. The red dot-dashed line shows the result of a system where electrons can only hop to nearest neighbors. The blue dashed line shows the fitting results. We can see that $G^\text{c}$ obeys the Mott's law $G^\text{c} \propto \exp\left[-(T_0/T)^{1/2}\right]$ in the VRH regime and a simple thermal activation behavior $G^\text{c} \propto e^{-T_0'/T}$ in the NNH regime, where $T_0$ and $T_0'$ are constants with the dimension of the temperature. In the VRH regime, $P$ can


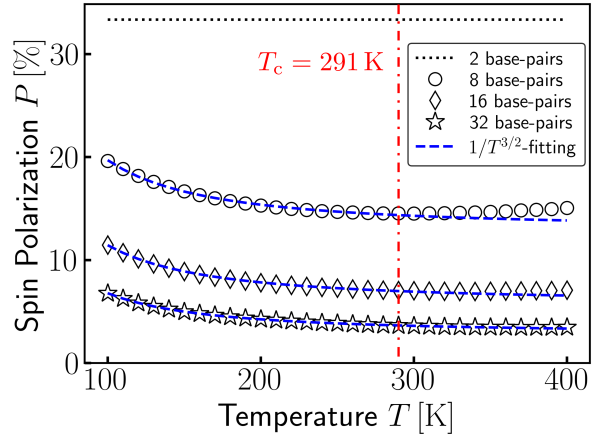

Supplementary Figure S3. Temperature dependences of the spin polarization without structural fluctuations for various molecular lengths.

be well fitted by $P \propto 1/T^{3/2}$ in the same manner. The deviations from the fitting result occur at the same crossover temperature $T_c = 291\,\mathrm{K}$ between $G^c$ and $P$. Therefore, we expect that the crossover of the mechanism of the electron transport from NNH to VRH observed in $G^c$ is closely related to the temperature dependence of $P$.

The molecular length dependence of the spin polarization can be also obtained by the same framework. In Fig. S3, we can see that $P$ shows the same power law behavior: $P \propto 1/T^{3/2}$ in each length except for the case of 2 base-pairs, where $P$ is given by $P_{2-\mathrm{bps}} = (G_{12}^{\uparrow\uparrow} - G_{12}^{\downarrow\downarrow})/(G_{12}^{\uparrow\uparrow} + G_{12}^{\downarrow\downarrow})$. Therefore, we expect that the temperature dependence of $P$ in the VRH regime is universal. On the other hand, our results suggest that the spin polarization decreases with increasing the length of the system. This seems to conflict with the general trends observed in various experiments [S.5]. The discrepancy between our results and the experimental ones may stem from the difference in the details of experimental conditions and the approximations such as the ignorance of overdamped phonons. These problems are left for our future study.

## SPATIAL PROFILE OF SPIN ACCUMULATION

In order to identify the origin of the temperature dependence of $P$, we have also investigated the spatial profile of the spin accumulation $\mu_i^s/eV = -V_i^s/V$. Fig. S4 shows the spatial profiles of $\mu_i^s/eV$ and their fitting results by $\sinh[(x - L/2)/l_\mathrm{sd}]$ for various temperatures. Here, $L = Na$ and $l_\mathrm{sd}$ are the system and the spin diffusion lengths.

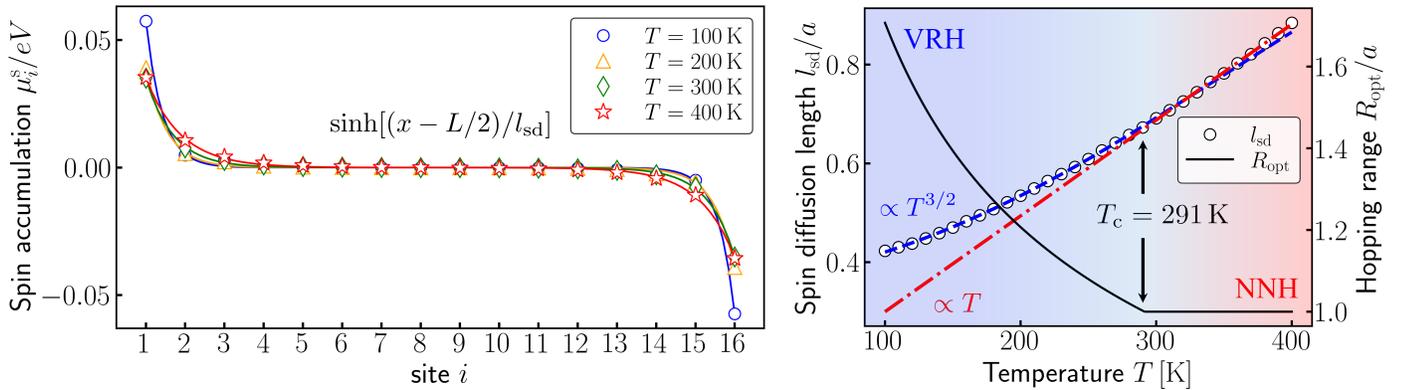

Supplementary Figure S4. (Left) The spatial profile of the spin accumulation $\mu_i^s$ for various temperatures with a molecular length of 16 base-pairs. The solid line shows the fitting results by $\sinh[(x - L/2)/l_\mathrm{sd}]$, where $L = Na$ and $l_\mathrm{sd}$ are the system and the spin diffusion lengths. (Right) The temperature dependence of the extracted spin diffusion length $l_\mathrm{sd}$ and the optimized hopping range $R_\mathrm{opt}$ in units of the lattice constant.

In the left panel of Fig. S4, we can see that each spin component accumulates in the same amount at the opposite ends of the system. At high temperatures, the decay length $l_{\rm sd}$ is longer and the spin accumulation is still visible within a few sites from both ends of the system, whereas at low temperatures, the amplitude of $\mu_i^{\rm s}$ at the two ends is larger than at high temperatures although it decays rapidly. The spatial profile of $\mu_i^{\rm s}$ is well fitted by $\sinh[(x - L/2)/l_{\rm sd}]$, indicating that the spin accumulation obeys a steady-state diffusion equation: $l_{\rm sd}^2 \partial_x^2 \mu^{\rm s}(x) = \mu^{\rm s}(x)$ with a temperature-dependent $l_{\rm sd}$.

Given that $P$ is proportional to $2\sinh[L/2l_{\rm sd}] \sim l_{\rm sd}^{-1}$, the temperature dependence of $P$ is governed by that of $l_{\rm sd}$. Therefore, we have further investigated the temperature dependence of $l_{\rm sd}$ and revealed that $l_{\rm sd}$ also shows a crossover from NNH to VRH at $T_{\rm c} = 291\,{\rm K}$, which is depicted in the right panel of Fig. S4. We have performed a temperature fitting and the result indicates that $l_{\rm sd} \propto T(a/R_{\rm opt})$, where the optimized hopping range $R_{\rm opt}$ is given by,

$$R_{\rm opt}(T) = \begin{cases} \xi \left(\dfrac{\Delta_\xi}{4k_{\rm B}T}\right)^{\frac{1}{d+1}} = a\left(\dfrac{T_{\rm c}}{T}\right)^{\frac{1}{d+1}} & T < T_{\rm c} \\ a & T > T_{\rm c} \end{cases}. \tag{S.48}$$

Thus, we can conclude that the temperature dependence of $P$ is originated from that of $R_{\rm opt}$ associated with the crossover of the underlying mechanism of electron transport from NNH to VRH.

---


\fi

\end{document}